\documentclass[12pt]{article}

\usepackage{a4,amsmath,amssymb,hyperref}

\allowdisplaybreaks[4]
\renewcommand{\title}[1]{\vspace{10mm}\noindent{\Large{\bf #1}}\vspace{8mm}}
\newcommand{\authors}[1]{\noindent{\large #1}\vspace{3mm}}
\newcommand{\address}[1]{{\itshape #1\vspace{2mm}}}

\makeatletter
\def\section{\@startsection{section}{1}{\z@}{-3.25ex plus -1ex minus
    -.2ex}{1.5ex plus .2ex}{\normalfont\large\bfseries}}
\def\subsection{\@startsection{subsection}{1}{\z@}{-3.25ex plus -1ex
    minus -.2ex}{1.5ex plus .2ex}{\normalfont\itshape}}

\renewenvironment{thebibliography}[1]
         {\section*{References}\frenchspacing\small
          \begin{list}{[\arabic{enumi}]}
         {\usecounter{enumi}\parsep=2pt\topsep 0pt
         \settowidth{\labelwidth}{[#1]}
         \leftmargin=\labelwidth\advance\leftmargin\labelsep
         \rightmargin=0pt\itemsep=0pt\sloppy}}{\end{list}}

\makeatother

\begin{document}

\begin{titlepage}

\hspace*{\fill}

\begin{center}

\title{8D-spectral triple on 4D-Moyal space and\\[2mm]
  the vacuum of noncommutative gauge theory} 

\authors{Harald {\sc Grosse}$^1$ and Raimar {\sc Wulkenhaar}$^2$}

\address{$^{1}$\,Institut f\"ur Theoretische Physik, Universit\"at Wien\\
Boltzmanngasse 5, A-1090 Wien, Austria}

\address{$^{2}$\,Mathematisches Institut der Westf\"alischen
  Wilhelms-Universit\"at\\
Einsteinstra\ss{}e 62, D-48149 M\"unster, Germany}

\footnotetext[1]{harald.grosse@univie.ac.at}
\footnotetext[2]{raimar@math.uni-muenster.de}

\vskip 3cm

\textbf{Abstract} \vskip 3mm 

\begin{minipage}{14cm}%
  Observing that the Hamiltonian of the renormalisable scalar field
  theory on 4-dimensional Moyal space $\mathcal{A}$ is the square of a
  Dirac operator $\mathcal{D}$ of spectral dimension 8, we complete
  $(\mathcal{A},\mathcal{D})$ to a compact 8-dimensional spectral
  triple. We add another
  Connes-Lott copy and compute the spectral action of the
  corresponding $U(1)$-Yang-Mills-Higgs model. We find that in the
  Higgs potential the square $\phi^2$ of the Higgs field is shifted to
  $\phi\star \phi + \mathrm {const} \cdot X_\mu \star X^\mu$, where
  $X_\mu$ is the covariant coordinate. The classical field equations
  of our model imply that the vacuum is no longer
  given by a constant Higgs field, but both the Higgs and gauge fields
  receive non-constant vacuum expectation values. 
\end{minipage}

\end{center}

\end{titlepage}

\setcounter{section}{-1}

\section{Preface}

In September 2007 we published the first version of this manuscript as
preprint arXiv:hep-th/0709.0095v1. We had found dimensional arguments
why previous attempts to construct a \emph{four-dimensional} spectral
triple for renormalisable scalar field theory on Moyal space with
harmonic oscillator potential \cite{Grosse:2004yu} had to fail. These
arguments showed the necessity of a doubling of the dimension, and
indeed we were able to identify an 8-dimensional Dirac operator
$\mathcal{D}$ which together with the Moyal algebra gave rise to a
reasonable spectral triple. We also computed the decisive part of the
resulting spectral action and completed it by gauge invariance.

In preparing a talk for the Oberwolfach meeting on ``Noncommutative
Geometry'' a few days later, one of us (R.W.) realised that the
dimensionality of our spectral triple is much more intricate
\cite{Wulkenhaar:2007??}. While $|\mathcal{D}|^{-8}$ is indeed of
Dixmier trace class, localising it with the operator $L_\star(f)$ of
left Moyal multiplication by a Schwartz function $f$ one has
$L_\star(f) |\mathcal{D}|^{-4}$ of Dixmier trace class. This dimension
drop was already visible in the different parts of the spectral
action. We thus concluded that the \emph{metric} dimension of our
spectral triple remains $d=4$ whereas the \emph{KO-dimension} is
$k=8$.

It became apparent that our spectral triple proposed in
arXiv:hep-th/0709.0095v1 was the shadow of a very rich mathematical
structure which had to be explored. Working out the details, the
corrected manuscript deviated more and more into a completely
different paper. Additionally, as the computation of the dimension
spectrum faced enormous difficulties, the commutative case was treated
first in \cite{Wulkenhaar:2009pv}. An important achievement of
\cite{Wulkenhaar:2009pv} was to understand that there are in fact two
Dirac operators $\mathcal{D}_1,\mathcal{D}_2$ which both relate to
supersymmetric quantum mechanics and which together permit a
realisation of the orientabity axiom. In arXiv:hep-th/0709.0095v1 we
had still pointed out that orientability cannot be recovered.

Eventually, all difficulties with the dimension spectrum in the Moyal
case, and several other mathematical issues, have been recently solved
in joint work of one of us with V.~Gayral \cite{Gayral:2011vu}. In
that paper the factorisation property of the Moyal
algebra\footnote{For any $f\in \mathcal{S}(\mathbb{R}^d)$ there are
  $f_1,f_2 \in \mathcal{S}(\mathbb{R}^d)$ with $f_1\star f_2=f$.} is
heavily used to prove rigourous $\mathcal{L}^p$-estimates for all
appearing operators. Using these H\"older type estimates a completely
different computation of the spectral action is given, which up to
typos confirms the result of arXiv:hep-th/0709.0095v1.

In summary, the paper \cite{Gayral:2011vu} supersedes
arXiv:hep-th/0709.0095v1 in all mathematical aspects. But
arXiv:hep-th/0709.0095v1 contains the precious heuristic discussion of
the dimensionality of the Dirac operator and a useful overview of
renormalisable field theories on Moyal space which both are lost in
\cite{Gayral:2011vu}. We therefore think that these parts of
arXiv:hep-th/0709.0095v1 and the original technique for computing the
spectral action are interesting enough to justify, in spite of four
years of delay, a corrected version. Although we know many things
better now, we limit ourselves to error corrections. In particular,
the historical introduction and the notation is unchanged. We silently
correct the typos in the spectral action as identified in
\cite[footnote 3]{Gayral:2011vu}. The original section about solutions
of the field equation (and comments in the introduction referring to
it) is completely removed. As pointed out to us by A.~Marcillaud de
Goursac, our formula for the Moyal product in radial coordinates was
wrong and with it our original conclusions. For a discussion of the
vacuum configuration of this type of action we refer to 
\cite{deGoursac:2008rb}.

\section{Introduction}

\subsection{Renormalisable field theories on Moyal space}

Renormalisable field theories on Moyal space are by now in mature
state. In the first renormalisation proof \cite{Grosse:2004yu}, the
matrix base of the Moyal plane was a central philosophy, because we
wanted to avoid convergence subtleties with the oscillating integrals
in momentum space. We traded the simple matrix product interaction in
for a complicated (but manifestly positive) propagator and used exact
renormalisation group equations to estimate the ribbon graphs. The
technically most challenging part was a brute-force analysis
\cite{Grosse:2003aj} of all possible contractions of ribbon graphs.
The scale analysis led to the existence of an additional marginal
coupling in the $\phi^4$-model, which corresponds to a harmonic
oscillator potential for the free field. Later on, we interpreted this
term as required by Langmann-Szabo duality \cite{Langmann:2002cc}.  A
summary of these ideas can be found in \cite{Grosse:2005da}.

The renormalisation proof was considerably simplified by switching to
multi-scale analysis as the renormalisation scheme. The first version
still relied on the matrix base \cite{Rivasseau:2005bh}. Once the
bounds for the sliced propagator being proven (which is tedious), one
obtains in an efficient way the power-counting theorem in terms 
of the topology of the graph. Subsequently, the renormalisation proof
was also achieved by multi-scale analysis in position space (which is
equivalent to momentum space by Langmann-Szabo duality)
\cite{Gurau:2005gd}, showing the equivalence of various
renormalisation schemes. Recently, the position space amplitude of an
arbitrary orientable graph was expressed as an integral over Symanzik
type hyperbolic polynomials \cite{Gurau:2006yc}. With all inner
integrations carried out, this is the most condensed way of writing
Feynman graph amplitudes. See also \cite{Rivasseau:2007qx} for the
more complicated case of ``critical'' models.

Additionally, we noticed that the $\beta$-function of the
renormalisable noncommutative $\phi^4$-model tends to zero at large
energy scales. This is opposite to the commutative case and supports
the hope that a non-perturbative construction of the model is within
reach \cite{Rivasseau:2007fr,Magnen:2007uy}. The one-loop
$\beta$-function was first computed in \cite{Grosse:2004by} (its
peculiar feature was noticed in \cite{Grosse:2005da}). Roughly
speaking, there is a one-loop wavefunction renormalisation in the
model (absent in the commutative case), which for large energy scales
exactly compensates the renormalisation of the four-point function.
Then, in \cite{Disertori:2006uy} it was shown that at the self-duality
point $\Omega=1$ (where $\Omega$ is the frequency of the harmonic
oscillator potential in natural units), the $\beta$-function vanishes
up to three-loop order. Eventually, in \cite{Disertori:2006nq} the
vanishing of the $\beta$-function (at $\Omega=1$) was proven to all
orders, which means that the Landau ghost is absent in noncommutative
$\phi^4_4$-theory: Wave function renormalisation exactly compensates
the renormalisation of the four-point function, so that the flow
between the bare and the renormalised coupling is bounded.  The main
tool in this proof is a clever combination of the Ward identity
relative to unitary transformations with the Schwinger-Dyson
equations. Strictly speaking, the proof requires $\Omega=1$, but using
the bounds established in \cite{Rivasseau:2005bh}, it is plausible
that the renormalisation flow of the coupling is bounded for $0<\Omega<1$,
too. 

A good review of these exciting developments is
\cite{Rivasseau:2007ab}. The relation to previous attempts to
renormalise noncommutative field theories is discussed in
\cite{Wulkenhaar:2006si}.

The  importance  of  the   self-duality  case  was  first  noticed  in
\cite{Langmann:2003cg,Langmann:2003if} where an exact non-perturbative
solution of a complex scalar field theory on Moyal space with critical
magnetic background field was  constructed. The UV-fixed point of this
model is trivial.  In \cite{Grosse:2005ig,Grosse:2006tc} a non-trivial
exactly solvable (and just  renormalisable) field theory was obtained,
the noncommutative $\phi^3_6$-model at the  self-duality point. Here,
self-duality  relates   this  model  to   the  Kontsevich-model.   For
$\phi^3_4$, see \cite{Grosse:2006qv}.

There is also considerable progress with other than scalar field
models on Moyal space. In
\cite{VignesTourneret:2006nb,VignesTourneret:2006xa} renormalisation
to all orders of the duality-covariant orientable Gross-Neveu model
was shown. To put it into context with the work we present here, it is
important to stress that the Dirac operator in
\cite{VignesTourneret:2006nb,VignesTourneret:2006xa} is \emph{not} the
square root of the harmonic oscillator Hamiltonian appearing in the
$\phi^4$-model of \cite{Grosse:2004yu} and following treatments. It is
precisely in this paper where we construct such a square root and
analyse its properties. The Dirac operator of the Gross-Neveu model is
of the type studied (for scalar fields) in
\cite{Langmann:2003cg,Langmann:2003if}, just describing the influence
of a constant magnetic background field. Its spectrum is very different
from the harmonic oscillator (there is e.g.\ infinite degeneracy).
This fact can also be seen from a different structure of the
propagator in position space \cite{Gurau:2005qm}, which made the
renormalisation of the Gross-Neveu model technically more difficult.
In some sense, the magnetic background field is not needed for
renormalisation of complex scalar fields, as already argued in
\cite{Chepelev:2000hm} (in the massive case a new counterterm is
generated, though). See \cite{Lakhoua:2007ra} for the one-loop
$\beta$-function of this model.

The most interesting field theories are Yang-Mills theories, which we
also would like to see in renormalisable form on Moyal space. Usual
Yang-Mills theory on Moyal space (without modifications of the action
by something similar to an oscillator potential) is known to be not
renormalisable \cite{Matusis:2000jf}. Yang-Mills theories in
noncommutative geometry \cite{Connes:1994yd} are
naturally obtained from the spectral action principle
\cite{Connes:1996gi,Chamseddine:1996zu} relative to an appropriate
Dirac operator. In this way, a beautiful reformulation of the standard
model of particle physics was obtained, see \cite{Chamseddine:2006ep}
for its most recent version. Moyal space with undeformed Dirac
operator is a (non-compact) spectral triple \cite{Gayral:2003dm}. The
corresponding spectral action was computed in \cite{Gayral:2004ww},
with the result that it is the usual Yang-Mills action on Moyal space
(which is not renormalisable).  The magnetic background field Dirac
operator of the Gross-Neveu model gives the same usual Yang-Mills
action, too.  

To obtain a gauge theory with sort of oscillator potential via the
spectral action principle, we need a Dirac operator with similar
spectrum as the square root of the harmonic oscillator.
Unfortunately, all attempts to produce such a Dirac operator failed so
far, and here we can report progress in this paper. As
workaround we translated the physical interpretation of the spectral
action (to describe a one-loop effective action of fermions in a
classical external gauge field) from fermions to scalar fields. In
\cite{Gayral:2004cs} this method was already worked out for general
(isospectral) Rieffel deformations \cite{Rieffel:1993??}. We finished
the computation almost simultaneously in position space \cite{de
  Goursac:2007gq} and in the matrix base \cite{Grosse:2007dm}. See
also \cite{Grosse:2006hh,Wallet:2007em}. As a result, there are two
additional terms to the Yang-Mills action, namely the integral over
$\tilde{X}_\mu\star \tilde{X}^\mu$ and over its square, where
$\tilde{X}_\mu(x)=(\Theta^{-1})_{\mu\nu} x^\nu + A_\mu(x)$ is a
covariant coordinate \cite{Madore:2000en}. The existence of such a
term was conjectured in \cite[p.\ 90]{raimar-habil}.

The problem with the effective action derived in \cite{de
  Goursac:2007gq,Grosse:2007dm} is that, expanding $\tilde{X}_\mu\star
\tilde{X}^\mu$ and its square, there is a linear term in the gauge
field $A_\mu$. The consequence is that $A_\mu=0$ \emph{is not a stable
solution of the classical field equation}. Any attempt to solve the
classical field equations resulting from \cite{de
  Goursac:2007gq,Grosse:2007dm} failed so far. To circumvent
the vacuum problem, in \cite{Blaschke:2007vc} an oscillator potential
for the gauge field was achieved solely from a generalised ghost
sector, in a BRST-invariant way. Although a one-loop calculation is
likely to produce the $\tilde{X}_\mu\star \tilde{X}^\mu$ terms as in
\cite{de Goursac:2007gq,Grosse:2007dm}, the investigations in
\cite{Blaschke:2007vc} demonstrate the enormous freedom of
constructing the ghost sector, which in some way will be needed to
obtain a manageable gauge field propagator. 

\subsection{Strategy of the paper}

Our paper starts from a simple observation, so simple that it is
embarrassing not having it earlier exploited. The harmonic oscillator
Hamiltonian $H$ in one-dimensional configuration space, thus
two-dimensional phase space, has spectrum $\omega(n+\frac{1}{2})$ with
$n\in \mathbb{N}$. Thus, $H^{-1}$ is a noncommutative infinitesimal
\cite{Connes:1996gi} of order one---the configuration space dimension.
The Hamiltonian $H$ generalises the Laplacian. The central object in
noncommutative geometry is the Dirac operator, which is a
(generalised) square root of the Laplacian. Now,
$\mathcal{D}=H^{\frac{1}{2}}$ is a noncommutative infinitesimal of
order one over two, two being the phase space dimension. Spectral
dimension is defined through the Dirac operator so that \emph{the
  spectral dimension of the harmonic oscillator is the phase space
  dimension}.

For field theory we are interested in four-dimensional Moyal
configuration space. The isospectral deformation would be a
four-dimensional spectral triple \cite{Gayral:2003dm}. But for
renormalisation of the $\phi^4_4$-theory we must promote the 4D
Laplace operator $-\Delta$ to the 4D harmonic oscillator Hamiltonian
$H=-\Delta + \Omega^2 \|x\|^2$.  According to the previous discussion,
the noncommutative dimension of the 4D harmonic oscillator Hamiltonian
is the phase space dimension, which is EIGHT, not four. We thus
understand why all attempts to find a 4D Dirac operator for the 4D
harmonic oscillator Hamiltonian necessarily failed. On the other hand,
it is absolutely trivial to write down an 8D Dirac operator so that
its square equals (up to a constant matrix) the 4D harmonic oscillator
Hamiltonian. This is what we do in Section 2. Additionally, we show
that our 8D-Dirac operator on 4D-Moyal space almost extends to an
eight-dimensional spectral triple in the original sense
\cite{Connes:1996gi}. The orientability axiom is violated. We do not
check Poincar\'e duality.

It is worthwhile to mention that the distinction between configuration
space and phase space dimension was crucial for the quantum field
theory on projective modules over the noncommutative torus
investigated in \cite{Gayral:2006wu}. There, $\mathbb{R}^2$ and the
2-dimensional space of holomorphic $\mathbb{C}^2$-function where
considered as projective modules, i.e.\ configuration space, over the
4D-noncommutative torus (which extends to a four-dimensional spectral
triple). The resulting Hamiltonian was precisely that of the
$2D$-harmonic oscillator, where the oscillator potential is naturally
obtained from the isospectral Dirac operator of the 4D-noncommutative
torus.  The field theory on 2D-configuration space was shown to be
one-loop renormalisable like a 4D-scalar field theory, four being the
phase space dimension of the noncommutative torus. The dimensional
relations with Moyal space were discussed to some extent in
\cite{Gayral:2006wu}. It was noticed that the heat kernel traces split
into a local integral over field monomials times a \emph{partial trace
  only} of the propagator (see also \cite{Gayral:2004cs}).  But the
true dimensionality of the harmonic oscillator Moyal space was not
realised.

Having the 8D-Dirac operator with harmonic oscillator spectrum, we
perform the standard procedure \cite{Connes:1996gi,Chamseddine:1996zu}
of noncommutative geometry to get to the spectral action. To make it a
little more interesting, we add in Section 3 another Connes-Lott copy
\cite{Connes:1990qp} and compute in Section 4 the spectral action for
the resulting two $U(1)$-Moyal Yang-Mills fields unified with a
complex Higgs field to a single noncommutative gauge field. This
extends the computation of \cite{de Goursac:2007gq,Grosse:2007dm}
where the effective scalar field action was (unfortunately) not
considered. It turns out that only the inclusion of the Higgs field
provides an understanding of the $\tilde{X}_\mu\star \tilde{X}^\mu$
terms: We find that they appear together with the Higgs field $\phi$
in a potential of the form $ (\alpha \tilde{X}_\mu\star \tilde{X}^\mu
+ \beta \phi\star \phi -1)^2$, for some positive numbers
$\alpha,\beta$. Thus, \emph{the origin of the non-trivial gauge field vacuum
is nothing but the standard Higgs mechanism}. We experience here a
further level of the unification of Higgs and gauge fields through
noncommutative geometry: Almost-commutative geometry obtained the
potential of the Higgs field as part of the unified Yang-Mills action.
Spatial noncommutativity intertwines gauge and Higgs field even
further so that the potential combines Higgs and gauge field on an
equal footing.

\section{A spectral triple in dimension 8}

The renormalisable real $\phi^4$-model on the $4$-dimensional Moyal
plane is characterised by the appearance of the
harmonic oscillator Hamiltonian
\begin{align}
H_m=-\frac{\partial^2}{\partial x_\mu \partial x^\mu} +\Omega^2 \tilde{x}^\mu
\tilde{x}_\mu +m^2
\end{align}
in the action functional \cite{Grosse:2004yu}, where
$\tilde{x}_\mu:=2(\Theta^{-1})_{\mu\nu} x^\nu$. For simplicity we
choose
\begin{align}
\Theta=\left(\begin{array}{cccc}
0 & \theta & 0 & 0 \\
-\theta & 0 & 0 & 0 \\
0 & 0 & 0 & \theta \\
0 & 0 & -\theta & 0
\end{array}\right)=:i\theta \sigma\;,\qquad \theta \in \mathbb{R}\;,
\end{align}
where $\sigma=\sigma_2\otimes 1_2$ consists of two copies of the
second Pauli matrix. We have $\Theta^{-1}=\frac{-i}{\theta} \sigma$.
It is then a well-known fact from quantum mechanics that the Hilbert
space $L^2(\mathbb{R}^4)$ has an orthonormal basis $\{
\psi_{\underline{n}} \}_{\underline{n} \in \mathbb{N}^4}$ of
eigenfunctions of $H_m$ with 
\[
H_m \psi_{\underline{n}}
=\frac{4\Omega}{\theta}\big( |{\underline{n}}| +2 
+ \tfrac{\theta m^2}{4\Omega}\big) 
 \psi_{\underline{n}}\;,\qquad |{\underline{n}}|=n_1+n_2+n_3+n_4~\text{ for } 
~{\underline{n}}=(n_1,n_2,n_3,n_4)\;.
\]
The inverse $H_m^{-1}$ extends to a selfadjoint compact operator on
$L^2(\mathbb{R}^4)$ with eigenvalues
\begin{align}
\lambda_n(m)=\Big(\frac{4\Omega}{\theta}\big( n+2 + \tfrac{\theta
  m^2}{4\Omega}\big)\Big)^{-1}\;,\qquad n \in \mathbb{N}\;.
\end{align}
The $n^{\mathrm{th}}$ eigenspace $E_n$ has dimension 
$\mathrm{dim} (E_n)= \binom{n+3}{3}$,
which is the number of possibilities to write $n$ as a sum of four ordered
natural numbers. This means that for $s>4$, the trace 
\begin{align}
\mathrm{Tr}(H_m^{-s}) = \frac{1}{6} \sum_{n=0}^\infty (n+3)(n+2)(n+1) 
(\lambda_n(m))^s
\end{align}
exists. The critical value $s=4$ characterises $H^{-4}$ as belonging
to the Dixmier trace ideal
$\mathcal{L}^{(1,\infty)}(L^2(\mathbb{R}^4))$ of compact operators
\cite{Connes:1994yd}. 

At first sight, $H^{-4} \in
\mathcal{L}^{(1,\infty)}(L^2(\mathbb{R}^4))$ seems to be related to
the four dimensional Moyal space under consideration.  However, recall
that in noncommutative geometry it is the \emph{Dirac operator} which
defines the dimension \cite{Connes:1996gi}. In a $d$-dimensional space
we require $|\mathcal{D}|^{-d} \in
\mathcal{L}^{(1,\infty)}(L^2(\mathbb{R}^4))$.  Identifying
$H=|\mathcal{D}|^2$, we notice the surprising fact that the
$4$-dimensional Moyal space has actually spectral dimension EIGHT.

In eight dimensions it is very easy to write down an appropriate Dirac
operator,
\begin{align}
\mathcal{D}_8= i\Gamma^\mu \partial_\mu + \Omega \Gamma^{\mu+4}
\tilde{x}_\mu\;. 
\end{align}
Here, the $\Gamma_k \in M_{16}(\mathbb{C})$, $k=1,\dots,8$ are the generators
of the 8-dimensional real Clifford algebra, satisfying
\begin{align}
\Gamma_k\Gamma_l + \Gamma_l \Gamma_k = 2\delta_{kl} 1 \;.
\end{align}
We agree
that latin indices run from 1 to 8 and greek indices from 1 to 4.  
Summation over repeated upper and lower indices is self-understood.

Accordingly, we take the Hilbert space
$\mathcal{H}_8=L^2(\mathbb{R}^4,\mathcal{S})$ of square integrable spinors
over FOUR-dimensional euclidean space, where the spinor bundle
has typical fibre $\mathbb{C}^{16}$. For $\psi \in \mathcal{H}_8$ we obtain
\begin{align}
\mathcal{D}_8^2 \psi = \big( (-\Delta + \Omega^2 \tilde{x}_\mu 
\tilde{x}^\mu )1 
+ \Sigma  \big) \psi\;,\qquad \Sigma := -i\Omega
(\Theta^{-1})_{\mu\nu} [\Gamma^\mu,\Gamma^{\nu+4}]\;,
\end{align}
with $\Delta=\partial_\mu \partial^\mu$.
Assuming a choice of the Clifford algebra where $\Sigma$ is diagonal, we
obtain up to the 16-fold multiplicity of each level and an unimportant shift in
the mass  exactly the spectrum of the harmonic oscillator Hamiltonian
$H$. In particular, $|\mathcal{D}_8|^{-8}$ belongs as required to the Dixmier
trace ideal $\mathcal{L}^{(1,\infty)}(L^2(\mathbb{R}^4,\mathcal{S}))$. 

As algebra $\mathcal{A}_8$ we take the unitalised Moyal
algebra\footnote{This choice of the algebra cannot verify the 
 orientability axiom in any form, because we cannot represent  
 the partition of unity localised at infinity (which
  belongs to $\mathcal{A}_8$) by derivatives of elements of the
  algebra (which is not possible with
  $\mathcal{A}_8$). This can be achieved by 
  an appropriate subalgebra of the multiplier algebra of
  $\mathbb{R}^4_\Theta$, see \cite{Gayral:2003dm}. But the
  orientability axiom fails anyway, so it suffices
  to work with $\mathcal{A}_8$.}
\begin{align}
\mathcal{A}_8 = \mathbb{R}^4_\Theta \oplus \mathbb{C}\;,
\end{align}
where $\mathbb{R}^4_\Theta$ is as a vector space given by the Schwarz
class functions on $\mathbb{R}^4$, equipped with the Moyal product
\begin{align}
(f\star g)(x) = \int d^4 y\,\frac{d^4 k}{(2\pi)^4}  \,f(x{+}\tfrac{1}{2}
\Theta \cdot k) \,g(x{+}y)\, \mathrm{e}^{i\langle k,y\rangle}\;,
\qquad f,g \in \mathcal{A}_8 \;.
\label{Moyal}
\end{align}
The Moyal product extends to constant functions using the integral
representation of the Dirac distribution.

The algebra $\mathcal{A}_8$ acts on $\mathcal{H}_8$ also by
componentwise Moyal product, $\star: \mathcal{A}_8 \times
\mathcal{H}_8 \to \mathcal{H}_8$ (we refer to \cite{Gayral:2003dm} for
the necessary extension of the Moyal product). Clearly, the smooth
spinors form a finitely generated projective module over
$\mathcal{A}_8$.

We compute the commutator of that action with the Dirac operator,
taking for smooth spinors the identity $2x^\mu \psi=x\star \psi + \psi
\star x$ into account, as well as the relation $[x^\nu ,f]_\star
=i\Theta^{\nu\rho} \partial_\rho f$:
\begin{align}
&\mathcal{D}_8 (f \star  \psi)- f\star (\mathcal{D}_8 \psi)
\nonumber
\\
&=  i\Gamma^\mu ((\partial_\mu f) \star  \psi 
+ f \star \partial_\mu \psi)
+ \tfrac{1}{2} \Omega \Gamma^{\mu+4}
(\tilde{x}_\mu \star (f\star \psi) + 
(f\star \psi) \star\tilde{x}_\mu )
\nonumber
\\
&-  i\Gamma^\mu f \star \partial_\mu \psi
- \tfrac{1}{2} \Omega \Gamma^{\mu+4}
(f\star (\tilde{x}_\mu \star \psi) +
f\star (\psi \star\tilde{x}_\mu ))
\nonumber
\\
&= \big( i(\Gamma^\mu + \Omega \Gamma^{\mu+4})
(\partial_\mu f) \big) \star  \psi \;.
\end{align}
Thus, just the four-dimensional differential of $f$ appears, no
$x$-multiplication! This differential is represented on $\mathcal
{H}_8$ by $\pi(dx^\mu)= \Gamma^\mu + \Omega \Gamma^{\mu+4}$, and it is
bounded. It commutes with Moyal multiplication from the right, so that
the order-one condition is achieved in the usual way.  However, the
algebra generated by $[\mathcal{D}_8,\mathcal{A}_8]$ and
$\mathcal{A}_8$ does not contain the chirality matrix $\Gamma_9$ so
that the orientability axiom does not hold. The ingredients of the
spectral triple which just rely on the Clifford algebra (dimension
table) are automatically satisfied. We do not check Poincar\'e
duality. In conclusion, up to the orientability axiom (and possibly
Poincar\'e duality), $(\mathcal{A}_8,\mathcal{H}_8,\mathcal{D}_8)$
forms a spectral triple of dimension 8.

\section{$U(1)$-Higgs model}

In the Connes-Lott spirit \cite{Connes:1990qp} we take the tensor
product of the 8-dimensional spectral triple
$(\mathcal{A}_8,\mathcal{H}_8,\mathcal{D}_8,\Gamma_9)$ with the finite
Higgs spectral triple $(\mathbb{C}\oplus\mathbb{C}, \mathbb{C}^2,
M\sigma_1)$. The Dirac operator $\mathcal{D}=\mathcal{D}_8 \otimes 1 +
\Gamma_9 \otimes M\sigma_1$ of the product triple becomes
\begin{align}
\mathcal{D} =\left(\begin{array}{cc} 
\mathcal{D}_8 & M \Gamma_9 \\ 
M \Gamma_9  & \mathcal{D}_8
\end{array}\right)\;.
\end{align}
In this representation, the algebra is $\mathcal{A}_8 \oplus \mathcal{A}_8
\ni (f,g)$, which acts on $\mathcal{H}=\mathcal{H}_8 \oplus \mathcal{H}_8$ by
diagonal Moyal multiplication. The commutator of $\mathcal{D}$ with $(f,g)$ is
\begin{align}
[\mathcal{D},(f,g)] =\left(\begin{array}{cc} 
i(\Gamma^\mu + \Omega \Gamma^{\mu+4})
L_\star(\partial_\mu f)  & M \Gamma_9 L_\star(g-f)\\ 
M \Gamma_9  L_\star (f-g) & i(\Gamma^\mu + \Omega \Gamma^{\mu+4})
L_\star(\partial_\mu g)
\end{array}\right)\;,
\end{align}
where $L_\star(f)\psi=f\star \psi$ is left Moyal multiplication.
This shows that selfadjoint fluctuated Dirac operators $\mathcal{D}_A=
\mathcal{D} + \sum_i a_i [\mathcal{D},b_i]$ are of the
form 
\begin{align}
\mathcal{D}_A  =\left(\begin{array}{cc} 
\mathcal{D}_8 + (\Gamma^\mu + \Omega \Gamma^{\mu+4})L_\star(A_\mu)
  & \Gamma_9 L_\star({\phi}) \\ \Gamma_9 L_\star(\bar{\phi})  & 
\mathcal{D}_8 + (\Gamma^\mu + \Omega \Gamma^{\mu+4}) L_\star(B_\mu)
\end{array}\right)\;,
\end{align}
for real fields $A_\mu,B_\mu \in \mathcal{A}_8$ and a complex field 
$\phi \in \mathcal{A}_8$. The square of $\mathcal{D}_A$ is
\begin{align}
\mathcal{D}_A^2 &= \left(\begin{array}{cc} 
(H_0^2+ L_\star(\phi\star \bar{\phi})) 1 +\Sigma  +F_A   & 
i(\Gamma^\mu + \Omega \Gamma^{\mu+4})\Gamma_9 L_\star(D_\mu \phi) 
\\[1ex]
i (\Gamma^\mu + \Omega \Gamma^{\mu+4})\Gamma_9
L_\star(\overline{D_\mu \phi}) & (H_0^2+ L_\star(\bar{\phi}\star \phi))1 
+\Sigma  +F_B
\end{array}\right)  \;,
\label{DA2}
\end{align}
where 
\begin{align} 
D_\mu \phi &:= \partial_\mu \phi -iA\star \phi+i\phi \star B\;,
\\[1ex]
F_A &:= \{\mathcal{D}_8, (\Gamma^\mu + \Omega \Gamma^{\mu+4}) L_\star(A_\mu)\} 
+ (\Gamma^\mu + \Omega \Gamma^{\mu+4})
(\Gamma^\nu + \Omega \Gamma^{\nu+4})
L_\star(A_\mu \star A_\nu)
\nonumber
\\
&=  \big\{ L_\star(A^\mu) ,  i \partial_\mu + \Omega^2 
M_\bullet(\tilde{x}_\mu) \big\}
+ (1+\Omega^2) L_\star(A_\mu \star A^\mu) \nonumber
\\
& 
+ i \big(\tfrac{1}{4}
  [\Gamma^\mu,\Gamma^\nu] +\tfrac{1}{4} \Omega^2
  [\Gamma^{\mu+4},\Gamma^{\nu+4}] + \Omega \Gamma^\mu \Gamma^{\nu+4} \big)
  L_\star(F^A_{\mu\nu})\;,
\end{align}
and similarly for $F_B$. In this expression, 
$F^A_{\mu\nu} = \partial_\mu A_\nu - \partial_\nu A_\mu -i(A_\mu \star
A_\nu - A_\nu \star A_\mu)$ is the field strength and
$(M_\bullet(\tilde{x}_\mu) \psi)(x) = \tilde{x}_\mu \psi(x)$ is ordinary
local multiplication.

\section{The spectral action}

\subsection{General remarks}

According to the spectral action principle
\cite{Connes:1996gi,Chamseddine:1996zu}, the bosonic action depends
only on the spectrum of the Dirac operator. Thus, by functional
calculus, the most general form of the bosonic action is
\begin{align}
S(\mathcal{D}_A)=\mathrm{Tr}\big(\chi(\mathcal{D}_A^2)\big)\;,
\end{align}
for some function $\chi:\mathbb{R}_+ \to \mathbb{R}_+$ for which the
Hilbert space trace exists. By Laplace transformation one has
\begin{align}
S(\mathcal{D}_A)=\int_0^\infty dt\; 
\mathrm{Tr}(e^{-t \mathcal{D}_A^2}) \hat{\chi}(t)\;,  
\end{align}
where $\hat{\chi}$ is the (inverse) Laplace transform of $\chi$, i.e.\ 
$\chi(s)=\int_0^\infty dt\;e^{-st} \hat{\chi}(t)$. Assuming the heat
kernel has an asymptotic expansion
\begin{align}
e^{-t \mathcal{D}_A^2} = \sum_{z=-\delta}^\infty a_z(\mathcal{D}_A^2) 
t^{z}\;,\qquad \delta \in \mathbb{N}\;,
\end{align}
we obtain
\begin{align}
S(\mathcal{D}_A)= \sum_{z=-\delta}^\infty \mathrm{Tr}(a_z(\mathcal{D}_A^2)) 
\int_0^\infty dt\; t^{z} \hat{\chi}(t)
=: \sum_{z=-\delta}^\infty \chi_z\,\mathrm{Tr}(a_z(\mathcal{D}_A^2)) \;.
\end{align}
For compact manifolds, the most singular order $\delta$ is half of the
dimension according to Weyl's theorem. To compute the $\chi_z$ we have
to distinguish the cases $z \in \mathbb{N}$ and $z \notin \mathbb{N}$.
First, 
\begin{align}
\int_0^\infty ds\;s^{-z-1} \chi(s) 
&=\int_0^\infty ds \int_0^\infty dt\;e^{-st} s^{-z-1} \hat{\chi}(t) 
=\Gamma(-z) \int_0^\infty dt\; t^{z} \hat{\chi}(t) \;,
\end{align}
which yields the coefficients $\chi_{z}$ unless $z \in \mathbb{N}$.
For $z=k\in \mathbb{N}$ we have instead 
\begin{align}
\int_0^\infty \!\!dt\; t^{k} \hat{\chi}(t)
&=\lim_{s\to 0} \int_0^\infty \!\! dt\; e^{-st} t^{k} \hat{\chi}(t)
=\lim_{s\to 0}\; (-1)^{k} \frac{\partial^{k}}{\partial
  s^{k}}  \int_0^\infty dt\; e^{-st} \hat{\chi}(t)
\nonumber
\\*
&=\lim_{s\to 0} \,(-1)^{k} \frac{\partial^{k} \chi}{\partial s^{k}}(s)
=(-1)^{k}   \chi^{(k)}(0)\;.
\end{align}
In summary,
\begin{subequations}
\begin{align}
\chi_{z} &= \frac{1}{\Gamma(-z)}\int_0^\infty
ds\;s^{-z-1} \chi(s) && \text{for } z \notin \mathbb{N}\;,
\\
\chi_{k} &= (-1)^{k}  \chi^{(k)}(0) && \text{for } k \in \mathbb{N} \;.
\end{align}
\end{subequations}

In a position space basis, the Hilbert space trace is given by 
\begin{align}
\mathrm{Tr}(e^{-t \mathcal{D}_A^2})=\int_{\mathbb{R}^4} dx \;
\mathrm{tr}\big((e^{-t \mathcal{D}_A^2})(x,x)\big)\;,
\end{align}
where $\mathrm{tr}$ denotes the matrix trace (including the Clifford
algebra) and $(e^{-t \mathcal{D}_A^2})(x,y)$ is the heat kernel.
To obtain the heat kernel coefficients $a_z(\mathcal{D}_A^2)$, we write
\begin{align}
\mathcal{D}_{A=0}^2:=\mathrm{H}_0\;,\qquad \mathcal{D}_A^2=:
\mathrm{H}_0-V\;,
\end{align}
and consider the Duhamel expansion (see \cite{Gayral:2005ih} for more
information) 
\begin{align}
e^{-t_0 (\mathrm{H}_0-V)} 
&=
 e^{-t_0 \mathrm{H}_0} - \int_0^{t_0}  \!\!dt_1 \;
\frac{d}{dt_1} \big(e^{-(t_0-t_1) (\mathrm{H}_0-V)} 
e^{-t_1 \mathrm{H}_0}\big) \nonumber 
\\
&=
 e^{-t_0 \mathrm{H}_0} + \int_0^{t_0}  \!\!dt_1 \; 
\big(e^{-(t_0-t_1) (\mathrm{H}_0-V)} 
V e^{-t_1 \mathrm{H}_0}\big) \nonumber 
\\
&= e^{-t_0 \mathrm{H}_0} + \int_0^{t_0} \!\! dt_1 \; 
\big(e^{-(t_0-t_1) \mathrm{H}_0} 
V e^{-t_1 \mathrm{H}_0}\big) \nonumber
\\
& + \int_0^{t_0} \!\! dt_1 \int_0^{t_0-t_1} \!\! dt_2 
\; \big(e^{-(t_0-t_1-t_2) \mathrm{H}_0} 
V e^{-t_2 \mathrm{H}_0}V e^{-t_1 \mathrm{H}_0}
\big) +\dots \nonumber 
\\
& + \int_0^{t_0} \!\! dt_1 \dots 
\int_0^{t_0-t_1-\dots-t_{n-1}} \!\! dt_n 
\; \big(e^{-(t_0-t_1-\dots-t_n) \mathrm{H}_0} 
(V e^{-t_n \mathrm{H}_0})\cdots (V e^{-t_1 \mathrm{H}_0})
\big) 
+\dots \nonumber
\\
&= e^{-t_0 \mathrm{H}_0} + \sum_{n=1}^\infty t_0^n \int_{\Delta^n} d^n\alpha
\Big(e^{-t_0(1-|\alpha|)\mathrm{H}_0}
\prod_{j=1}^n (V e^{-t_0\alpha_j \mathrm{H}_0})\Big) \;,
\label{Duhamel}
\end{align}
where the integration is performed over the standard $n$-simplex
$\Delta^n :=\{\alpha:=(\alpha_1,\dots,\alpha_n) \in \mathbb{R}^n\;,~
\alpha_i \geq 0\;,~ |\alpha|:=\alpha_1+\dots+\alpha_n \leq 1\}$.

\subsection{Position space kernels}

According to (\ref{DA2}) we have $\mathrm{H}_0=H_01_{32} + \Sigma
1_2$. Its position space kernel is
\begin{align}
(e^{-t \mathrm{H}_0})(x,y)
&= \int d^4z \;(e^{-t H_0 1_{32}})(x,z) \;(e^{-t \Sigma 1_2})(z,y)
= e^{-t \Sigma 1_2} (e^{-t H_0})(x,y)\nonumber
\\ 
&= \Big(\frac{\tilde{\Omega}}{2\pi\sinh(2\tilde{\Omega} t)}
\Big)^2
e^{-t\Sigma 1_2 -\frac{\tilde{\Omega}}{4}\big(
\coth(\tilde{\Omega}t)|x-y|^2 
+\tanh(\tilde{\Omega}t) |x+y|^2 \big)}\;,
\end{align}
where the main part is given by the four-dimensional Mehler kernel
(see e.g.\ \cite{Simon:1979??}), with
$\tilde{\Omega}:=\frac{2\Omega}{\theta}$ and $|x|^2:=x_\mu x^\mu$. It
will be convenient to distinguish the following vertices in
(\ref{DA2}):
\begin{subequations}
\begin{align}
V_\phi &=  - L_\star(\phi \star \phi) 1_{16}\;, \\
V_{D\phi} &=  -i L_\star(D_\mu \phi) (\Gamma^\mu+\Omega
\Gamma^{\mu+4})\Gamma_9  \;,
\\
V_A &= 
- (1+\Omega^2) L_\star(A_\mu \star A^\mu) \;,
\\
V_{DA} &= -\big\{ L_\star(A^\mu) ,  i \partial_\mu + \Omega^2 
M_\bullet(\tilde{x}_\mu) \big\}\;,
\\
V_{FA} &=-
i L_\star(F^A_{\mu\nu})
\big(\tfrac{1}{4}
  [\Gamma^\mu,\Gamma^\nu] +\tfrac{\Omega^2}{4}
  [\Gamma^{\mu+4},\Gamma^{\nu+4}] 
+ \tfrac{\Omega}{2} \Gamma^\mu \Gamma^{\nu+4} 
- \tfrac{\Omega}{2} \Gamma^\nu \Gamma^{\mu+4} 
\big)
  \;,
\end{align}
\end{subequations}
and similarly for $V_B$, $V_{DB}$ and $V_{FB}$. 

We compute the necessary position space kernels: 
\begin{align}
(L_\star (f) g)(x) 
&= \int d^4 y\,\Big( \int\frac{d^4 k}{(2\pi)^4}  \,f(x{+}\tfrac{1}{2}
\Theta \cdot k) \,\mathrm{e}^{i\langle k,y-x\rangle}\Big) g(y)
\nonumber
\\
&=   \int d^4 y\,\Big( \frac{1}{\pi^4\theta^4} \int d^4 z  \,f(z)
\,\mathrm{e}^{2i(\langle x,\Theta^{-1} y\rangle
+\langle y,\Theta^{-1} z\rangle
+\langle z,\Theta^{-1} x\rangle)}\Big) g(y)\;,
\end{align}
from which we get 
\begin{align}
(L_\star (f))(x,y)=\frac{1}{\pi^4\theta^4} \int d^4 z  \,f(z)
\,\mathrm{e}^{i\langle x-y,\Theta^{-1} (x+y)\rangle
+2i\langle z,\Theta^{-1} (x-y)\rangle}\;.
\end{align}
Next, we compute 
\begin{align}
&\big(\big\{ L_\star(A^\mu) , i \partial_\mu + \Omega^2 
M_\bullet(\tilde{x}_\mu) \big\} g\big)(x) 
\nonumber
\\
&=   \int d^4 y\,\Big( \frac{1}{\pi^4\theta^4} \int d^4 z  \,A^\mu(z)
\,\mathrm{e}^{2i(\langle x,\Theta^{-1} y\rangle
+\langle y,\Theta^{-1} z\rangle
+\langle z,\Theta^{-1} x\rangle)}\Big) 
\Big(i \frac{\partial g}{\partial y^\mu }(y) 
+  \Omega^2  \tilde{y}_\mu g(y)\Big)
\nonumber
\\
&+ \Big(i\frac{\partial}{\partial x^\mu} +\Omega^2 \tilde{x}_\mu\Big)
\Big( \int d^4 y\,\Big( \frac{1}{\pi^4\theta^4} \int d^4 z  \,A^\mu(z)
\,\mathrm{e}^{2i(\langle x,\Theta^{-1} y\rangle
+\langle y,\Theta^{-1} z\rangle
+\langle z,\Theta^{-1} x\rangle)}\Big) g(y)\Big)
\nonumber
\\
&=   \int \!\! d^4 y\,\Big( \frac{1}{\pi^4\theta^4} \int \!\! d^4z  \,
(2\tilde{z}^\mu -(1{-}\Omega^2)(\tilde{x}^\mu{+}\tilde{y}^\mu))A_\mu(z) 
\,\mathrm{e}^{2i(\langle x,\Theta^{-1} y\rangle
+\langle y,\Theta^{-1} z\rangle
+\langle z,\Theta^{-1} x\rangle)}\Big) g(y)\;.
\end{align}
Therefore, the position space kernel of a Moyal-derivative vertex is 
\begin{align}
&\big\{ L_\star(A^\mu) , i \partial_\mu + \Omega^2 
M_\bullet(\tilde{x}_\mu) \big\}(x,y)
\nonumber
\\
&=  \frac{1}{\pi^4\theta^4} \int \!\! d^4z  \,
(2\tilde{z}^\mu -(1{-}\Omega^2)(\tilde{x}^\mu{+}\tilde{y}^\mu))A_\mu(z) 
\,\mathrm{e}^{2i(\langle x,\Theta^{-1} y\rangle
+\langle y,\Theta^{-1} z\rangle
+\langle z,\Theta^{-1} x\rangle)}\;.
\end{align}

\subsection{Computation of the traces}

The first term in the expansion (\ref{Duhamel}), which corresponds to
vacuum graphs, has the heat kernel expansion
\begin{align}
\mathrm{Tr}(e^{-t \mathrm{H}_0}) &=\mathrm{tr} \int d^4x \;
(e^{-t \mathrm{H}_0})(x,x)
=  \Big(\frac{\tilde{\Omega}}{2\pi\sinh(2\tilde{\Omega} t)} \Big)^2
2\,\mathrm{tr} \int d^4x \;
e^{-t\Sigma  -\tilde{\Omega}
\tanh(\tilde{\Omega}t) |x|^2} \nonumber
\\
&=  \frac{1}{8 \sinh^4(\tilde{\Omega} t)} 
\mathrm{tr} \big(e^{-t\Sigma}\big)\;.
\end{align}
We need the traces of the lowest powers of $\Sigma$:
\begin{align}
\mathrm{tr}(\Sigma^0)=16\;,\qquad  
\mathrm{tr}(\Sigma^2 )=16 \cdot \frac{16  \Omega^2}{\theta^2}\;,\qquad  
\mathrm{tr}(\Sigma^4)=16 \cdot \frac{640 \Omega^4}{\theta^4}\;.
\end{align}
All odd powers of $\Sigma$ are traceless. Therefore,
\begin{align}
\mathrm{Tr}(e^{-t \mathrm{H}_0}) &=
\frac{\theta^4}{8 \Omega^4 t^4}  + \frac{2\theta^2}{3 \Omega^2t^2}
+ \frac{52}{45} +\mathcal{O}(t^2)\;.
\end{align}
This reconfirms that the noncommutative space under consideration is
of dimension 8.

In the appendix we compute 
the first and second order $x$-$y$ integrals 
\begin{align}
& \int d^4x\, d^4y\; (e^{-tH_0})(y,x) V(x,y)\;,\\
& \int d^4x_1 \,d^4y_1\, d^4x_2 \,d^4y_2\; 
(e^{-(t-t_2)H_0})(y_2,x_1) V(x_1,y_1)
(e^{-t_2H_0})(y_1,x_2) V'(x_2,y_2)\;, \nonumber
\end{align}
where $V,V'$ stand for combinations of the Moyal and Moyal-derivative
vertices. In second order, we also perform a Taylor expansion about
coinciding external positions. It is remarkable that only terms of
order $t^{-1}$ and regular terms in $t$ appear, just as in
4D-Yang-Mills theory. Only the vacuum graphs behave like a
8D-model, for proper graphs only partial 4D-traces appear.

In the following, we only consider the trace of the $16$-dimensional
upper left corner containing the $A$-field and the structure
$\phi\star\bar{\phi}$. At the very end we add the lower right corner
where $A$ is replaced by $B$ and $\phi \leftrightarrow \bar{\phi}$.

With one $V_A$ or $V_{DA}$ vertex we see from (\ref{VLf}) and (\ref{VDA})
that the leading divergence after $t_1$-integration is $\sim
t^{-1}$. Therefore, the $\Sigma$ matrix gives no contribution up to
order $t^0$, so that the leading terms are 
\begin{align}
S_{(A+DA)}(t) & :=
\mathrm{Tr}\Big( \int_0^{t}  \!\!dt_1 \; \big(e^{-(t_0-t_1) (\mathrm{H}_0)} 
(V_A+V_{DA}) e^{-t_1 \mathrm{H}_0}\big)\Big)
\nonumber
\\
&= \frac{1}{\pi^2(1+\Omega^2)^2}
\int d^4 z \;\Big\{ 
-\frac{4\Omega^2}{(1+\Omega^2)} t^{-1}
\tilde{z}^\mu A_\mu(z)
+\frac{4\Omega^4}{(1+\Omega^2)^2} 
\tilde{z}^\mu A_\mu(z) \tilde{z}^\nu \tilde{z}_\nu
\nonumber
\\
& -  (1+\Omega^2) t^{-1} (A_\mu\star A^\mu)(z) 
+  \Omega^2 (A_\mu\star A^\mu)(z) 
\tilde{z}^\nu \tilde{z}_\nu \Big\} + \mathcal{O}(t)\;.
\end{align}
A single $V_{FA}$-vertex gets a non-vanishing trace of order $t^0$ 
together with one $\Sigma$-matrix, but the resulting integral 
$\int dz \;F_{\mu\nu}(z)$ vanishes. With two  $V_{FA}$-vertices and
the trace 
\begin{align}
&\mathrm{tr}\Big(
i \big(\tfrac{1}{4}
  [\Gamma^\mu,\Gamma^\nu] +\tfrac{1}{4} \Omega^2
  [\Gamma^{\mu+4},\Gamma^{\nu+4}] 
+ \tfrac{1}{2} \Omega \Gamma^\mu \Gamma^{\nu+4}
 - \tfrac{1}{2}\Omega \Gamma^\nu \Gamma^{\mu+4}  \big)  
\nonumber
\\
& \qquad \times i \big(\tfrac{1}{4}
  [\Gamma^\rho,\Gamma^\sigma] +\tfrac{1}{4} \Omega^2
  [\Gamma^{\rho+4},\Gamma^{\sigma+4}] 
+ \tfrac{1}{2} \Omega \Gamma^\rho \Gamma^{\sigma+4}
 - \tfrac{1}{2}\Omega \Gamma^\sigma \Gamma^{\rho+4}  \big)  
  \Big)
\nonumber
\\
& = 4(1+\Omega^2)^2 (\delta_{\mu\rho}\delta_{\nu\sigma}-
\delta_{\mu\sigma}\delta_{\nu\rho}) 
\end{align}
we find with (\ref{V2fg})
\begin{align}
S_{(FA)^2}(t) &:= \mathrm{Tr}\Big( \int_0^{t}  \!\!dt_1 
\int_0^{t-t_1}  \!\!dt_2 
\; e^{-(t_0-t_1-t_2) \mathrm{H}_0} 
V_{FA} e^{-t_2 \mathrm{H}_0}
V_{FA} e^{-t_1 \mathrm{H}_0}\Big)
\nonumber
\\
&= \frac{1}{4\pi^2} t^0
\int d^4z \; F^A_{\mu\nu}(z) F_A^{\mu\nu}(z)+\mathcal{O}(t)\;.
\end{align}

For two $V_{DA}$-vertices we obtain from (\ref{V2A0}) after some
integrations by parts
\begin{align}
S_{(DA)^2}(t) &:=\mathrm{Tr}\Big( \int_0^{t}  \!\!dt_1 
\int_0^{t-t_1}  \!\!dt_2 
\; e^{-(t_0-t_1-t_2) (\mathrm{H}_0)} 
V_{DA} e^{-t_2 \mathrm{H}_0}
V_{DA} e^{-t_1 \mathrm{H}_0}\Big)
\nonumber
\\
&= 16\int_0^{t}  \!\!dt_1 
\int_0^{t-t_1}  \!\!dt_2 
\; \frac{1}{(4\pi t)^2 (1+\Omega^2)^4} 
\int d^4 z 
\nonumber
\\
&
\times 
\Big( \frac{2(1-\Omega^2)^2(1+\Omega^2)}{t} A_\mu(z)A^\mu(z)
- 2\Omega^2 (1-\Omega^2)^2 
A_\mu(z)A^\mu(z) |\tilde{z}|^2
\nonumber
\\
&\qquad 
+ A^\mu(z)(\partial^\nu\partial_\nu  A_\mu)(z)
\Big(2(1-\Omega^2)^4 \frac{t_2(t-t_2)}{t^2} 
+2 \Omega^2(1-\Omega^2)^2 \Big) 
\nonumber
\\
& +16\Omega^4 \tilde{z}^\mu  A_\mu(z)\,
\tilde{z}^\nu A_\nu(z)
+ (1-\Omega^2)^4 \frac{t^2-4t_2t+4t_2^2}{t^2}
(\partial_\nu A_\mu)(z)\,(\partial^\mu A^\nu)(z)
\Big)
\nonumber
\\
&= \frac{1}{\pi^2 (1+\Omega^2)^2} 
\int d^4 z
\Big( \frac{(1-\Omega^2)^2}{1+\Omega^2} 
t^{-1} A_\mu \star A^\mu
\nonumber
\\*
&\qquad 
- \frac{(1-\Omega^2)^4}{6(1+\Omega^2)^2} \big(
(\partial^\nu A^\mu) \star (\partial_\nu  A_\mu)
- (\partial^\nu A^\mu) \star (\partial_\mu A_\nu)\big)
\nonumber
\\*
& \qquad - \frac{\Omega^2 (1-\Omega^2)^2}{(1+\Omega^2)^2} 
A_\mu \star A^\mu |\tilde{z}|^2
+\frac{8\Omega^4}{(1+\Omega^2)^2} 
\Big( (\tilde{z}\cdot A) \star (\tilde{z}\cdot A) \Big)(z)\;.
\end{align}
We have used 
\begin{align}
A^\mu(z) \tilde{z}_\nu \tilde{z}^\nu =  
A^\mu(z) \star (\tilde{z}_\nu \tilde{z}^\nu) 
+ i (\partial_\nu A^\mu)(z)\tilde{z}^\nu  
+ (\partial_\nu\partial^\nu A^\mu)(z) 
\end{align}
as well as $\int d^4z \;
A_\mu(z) (\partial_\nu A^\mu)(z)\,\tilde{z}^\nu=0$.

The $A$-linear and
$A$-bilinear part of the spectral action are given by  
the sum $S_{(A+DA)}+S_{(FA)^2}+S_{(DA)^2}$. As the spectral action is
manifestly gauge invariant, we simply complete the $A$-trilinear and 
$A$-quadrilinear terms in a gauge-invariant way. Introducing covariant
coordinates 
\begin{align}
\tilde{X}_{A}^\mu(z) :=\tfrac{\tilde{z}^\mu}{2}+A^\mu(z)\;,
\end{align}
with $\tilde{X}_{0}^\mu(z)= \tfrac{\tilde{z}^\mu}{2}$, 
we obtain the pure $A$-part of the spectral action to
\begin{align}
S_A(t) &= \frac{1}{\pi^2(1+\Omega^2)^2} 
\int d^4 z  \;\Big\{ - 
\frac{4\Omega^2}{1+\Omega^2}t^{-1}
\big( \tilde{X}^\mu_A  \star \tilde{X}_{A\mu}
- \tilde{X}^\mu_0  \star \tilde{X}_{0\mu}\big)
\nonumber
\\
& + \frac{t^0}{2} \Big(\frac{4\Omega^2}{1+\Omega^2}\Big)^2 
\big(\tilde{X}^\mu_A \star \tilde{X}_{A\mu}\star
\tilde{X}^\nu_A \star \tilde{X}_{A\nu}
- \tilde{X}^\mu_0 \star \tilde{X}_{0\mu}\star
\tilde{X}^\nu_0 \star \tilde{X}_{0\nu}\big)
\nonumber
\\
& + \Big(\frac{(1+\Omega^2)^2}{4}
- \frac{(1-\Omega^2)^2}{12 (1+\Omega^2)^2}\Big) 
t^0 \,F^A_{\mu\nu}\star  F_A^{\mu\nu}\Big\}(z)
+\mathcal{O}(t)\;.
\end{align}

The scalar field potential becomes
\begin{align}
S_{(\phi+\phi^2)}(t) 
&=\mathrm{Tr}\Big( \int_0^{t}  \!\!dt_1 \; \big(e^{-(t_0-t_1) \mathrm{H}_0} 
V_\phi e^{-t_1 \mathrm{H}_0}\big)\Big)
\nonumber
\\
&+ \mathrm{Tr}\Big( \int_0^{t}  \!\!dt_1 
\int_0^{t-t_1}  \!\!dt_2 
\; e^{-(t_0-t_1-t_2) \mathrm{H}_0} 
V_{\phi} e^{-t_2 \mathrm{H}_0}
V_{\phi} e^{-t_1 \mathrm{H}_0}\Big)
\nonumber
\\
&=  \frac{1}{\pi^2(1+\Omega^2)^2}
\int d^4z\;\Big( - t^{-1} 
\phi\star \bar{\phi} 
+\frac{\Omega^2|\tilde{z}|^2}{1+\Omega^2}
\phi\star \bar{\phi} 
+  \frac{1}{2} \phi\star \bar{\phi} \star 
\phi\star \bar{\phi}\Big)(z)\;.
\end{align}
The usual kinetic term of the scalar field comes from two
$D\phi$-vertices:
\begin{align}
S_{(D\phi)^2}(t) &= \mathrm{Tr}\Big( \int_0^{t}  \!\!dt_1 
\int_0^{t-t_1}  \!\!dt_2 
\; e^{-(t_0-t_1-t_2) \mathrm{H}_0} 
V_{D\phi} e^{-t_2 \mathrm{H}_0}
V_{\overline{D\phi}} e^{-t_1 \mathrm{H}_0}\Big)
\nonumber
\\
&=\frac{1}{2\pi^2(1+\Omega^2)} t^0 \int d^4z\; \big(D_\mu\phi\star 
\overline{D_\mu \phi}\big)(z) + \mathcal{O}(t)\;.
\end{align}
It remains the combination of $V_\phi$ with $V_A$ and $V_{DA}$, namely
$V_\phi V_A$, $V_{A} V_\phi$ as well as $V_{DA} V_\phi$, $V_\phi
V_{DA}$ and $V_\phi V_{DA} V_{DA}$, $V_{DA} V_\phi V_{DA}$, $V_{DA}
V_{DA} V_\phi$. At first order in $A$ we get from (\ref{VDAf})
\begin{align}
S_{(DA\phi+\phi DA)}(t)&= \mathrm{Tr}\Big( \int_0^{t}  \!\!dt_1 
\int_0^{t-t_1}  \!\!\!\!dt_2 
\; 
\Big(e^{-(t_0-t_1-t_2) \mathrm{H}_0} 
\Big(V_{\phi} e^{-t_2 \mathrm{H}_0}
V_{A} e^{-t_1 \mathrm{H}_0}
+
V_{A} e^{-t_2 \mathrm{H}_0}
V_{\phi} e^{-t_1 \mathrm{H}_0}\Big) \Big)
\nonumber
\\*
&= \frac{4\Omega^2}{\pi^2(1+\Omega^2)^3} 
\int d^4z\;\big(\phi\star \bar{\phi}\star (\tilde{z}^\mu
A_\mu)\big)(z)+\mathcal{O}(t)\;.
\end{align}
Completing the the $AA\phi\bar{\phi}$-term by gauge invariance, the
scalar field part of the spectral action becomes 
\begin{align}
S_\phi(t) = \frac{1}{\pi^2(1+\Omega^2)^2}&
\int d^4z\;  \Big( -t^{-1}\phi\star \bar{\phi} 
+\frac{1}{2} D_\mu\phi\star \overline{D_\mu \phi}
\nonumber
\\
& + \frac{1}{2} \phi\star\bar{\phi}\star \Big(
\phi\star\bar{\phi} +2\frac{4\Omega^2}{1+\Omega^2} 
\tilde{X}^\mu_A \star \tilde{X}_{A\mu}\Big)\Big)(z)\;.
\end{align}

To obtain the spectral action, we convert the Laplace-transform 
variable $t^{n}$ into $\chi_n$ and add the lower $B$-corner. The result
(including the vacuum contribution is 
\begin{align}
S&= \frac{\theta^4 \chi_{-4}}{8 \Omega^4}  + \frac{2\theta^2 \chi_{-2}
}{3\Omega^2} + \frac{52 \chi_0}{45} 
\nonumber
\\
& + \frac{\chi_0}{2\pi^2(1+\Omega^2)^2}
\int d^4z\;  \Big\{
\big(\tfrac{(1+\Omega^2)^2}{2}
- \tfrac{(1-\Omega^2)^4}{6 (1+\Omega^2)^2}\big) \,
(F^A_{\mu\nu}\star  F_A^{\mu\nu}+F^B_{\mu\nu}\star  F_B^{\mu\nu})
\nonumber
\\
& +\Big(\phi\star \bar{\phi} 
+ \tfrac{4\Omega^2}{1+\Omega^2} \tilde{X}^\mu_A \star 
\tilde{X}_{A\mu} -\frac{\chi_{-1}}{\chi_0}\Big)^2 
+\Big(\bar{\phi} \star \phi
+ \tfrac{4\Omega^2}{1+\Omega^2} \tilde{X}^\mu_B \star 
\tilde{X}_{B\mu} -\frac{\chi_{-1}}{\chi_0}\Big)^2 
\nonumber
\\
& 
-2  \Big(\tfrac{4\Omega^2}{1+\Omega^2}
\tilde{X}^\mu_0 \star \tilde{X}_{0\mu} -\frac{\chi_{-1}}{\chi_0}\Big)^2
+2(1+\Omega^2) D_\mu\phi\star \overline{D_\mu \phi}
\Big\}(z)  +\mathcal{O}(\chi_{1})\;.
\label{SpecAct}
\end{align}
The most important conclusion is that the squared covariant
derivatives combine with the Higgs field to a non-trivial
potential. This was not noticed in \cite{de
  Goursac:2007gq,Grosse:2007dm}.

\renewcommand{\thesection}{\Alph{section}}
\setcounter{section}{0}

\section{Appendix: Moyal integrals}

\allowdisplaybreaks[4]

\subsection{One Moyal vertex}

We compute a generic trace term with a change of variables 
$u=x-y$, $v=x+y$ with Jacobian $\frac{1}{16}$ (see \cite{Gurau:2005gd}):
\begin{align}
V_1(f)&:=\int d^4x \,d^4y\; (e^{-tH_0})(y,x) (L_\star(f))(x,y)
\nonumber
\\*
&= \frac{\tilde{\Omega}^2}{4\pi^2\sinh^2(2\tilde{\Omega} t)} 
\frac{1}{(2\pi\theta)^4} \int d^4 z  \,f(z)
\int d^4u\, d^4v\;
e^{-\frac{\tilde{\Omega}}{4}(\frac{ |u|^2}{\tanh(\tilde{\Omega}t)}
+\frac{|v|^2}{\coth(\tilde{\Omega}t)})
+i\langle u,\Theta^{-1} (v-2z)\rangle}
\nonumber
\\
&= \frac{1}{\cosh^4(\tilde{\Omega} t)} 
\frac{1}{(2\pi\theta)^4} \int d^4 z  \,f(z)
\int d^4v\;
e^{-\frac{\tilde{\Omega}\tanh(\tilde{\Omega}t)}{4}
(|v|^2+ \frac{4}{\tilde{\Omega}^2\theta^2}|v-2z|^2)}
\nonumber
\\
&= \frac{1}{\cosh^4(\tilde{\Omega} t)} 
\frac{1}{(2\pi\theta)^4} \int d^4 z  \,f(z)
\int d^4v\;
e^{-\frac{\tanh(\tilde{\Omega}t)}{2\theta \Omega}
((1+\Omega^2) |v-\frac{2}{1+\Omega^2}z|^2 +
\frac{4\Omega^2}{1+\Omega^2} |z|^2)}
\nonumber
\\
&= \frac{\tilde{\Omega}^2}{4\pi^2 (1+\Omega^2)^2 
\sinh^2(2\tilde{\Omega} t)} 
\int d^4 z  \,f(z)
e^{-\frac{\tilde{\Omega} \tanh(\tilde{\Omega}t)}{1+\Omega^2} |z|^2}\;.
\label{VLf}
\end{align}

\subsection{One Moyal+derivative vertex}

After a change of variables 
$u=x-y$, $v=x+y$ with Jacobian $\frac{1}{16}$, we have
\begin{align}
V_1(A) &:=\int d^4x d^4y\; (e^{-tH_0})(y,x) \big\{ L_\star(A^\mu) , 
i \partial_\mu + \Omega^2 M_\bullet(\tilde{x}_\mu) \big\}(x,y)
\nonumber
\\
&= \frac{\tilde{\Omega}^2}{4\pi^2\sinh^2(2\tilde{\Omega} t)} 
\frac{1}{(2\pi\theta)^4} 
\int d^4u \,d^4v \,d^4 z  \,A_\mu(z)(2\tilde{z}^\mu 
- (1-\Omega^2)\tilde{v}^\mu)
\nonumber
\\
& \times 
e^{-\frac{\tilde{\Omega}}{4}(\frac{ |u|^2}{\tanh(\tilde{\Omega}t)}
+\frac{|v|^2}{\coth(\tilde{\Omega}t)})
+i\langle u,\Theta^{-1} (v-2z)\rangle}
\nonumber
\\
&= \frac{\tilde{\Omega}^2}{4\pi^2\sinh^2(2\tilde{\Omega} t)} 
\frac{1}{(2\pi\theta)^4} 
\int d^4u \Big(\int d^4v \, d^4 z  \,A_\mu(z)\Big(2\tilde{z}^\mu 
+ 2i(1-\Omega^2)\frac{\partial}{\partial w_\mu}\Big)
\nonumber
\\
& \times 
e^{-\frac{\tilde{\Omega}}{4}(\frac{ |u|^2}{\tanh(\tilde{\Omega}t)}
+\frac{|v|^2}{\coth(\tilde{\Omega}t)})
+i\langle w,\Theta^{-1} v\rangle
-2i\langle u,\Theta^{-1} z\rangle}\Big|_{w=u}\Big)
\nonumber
\\
&= \frac{1}{\sinh^4(\tilde{\Omega} t)} 
\frac{1}{(2\pi\theta)^4} 
\int d^4u \Big(\int d^4 z  \,A_\mu(z)\Big(2\tilde{z}^\mu 
- \frac{2i(1-\Omega^2)}{\Omega \theta \tanh(\tilde{\Omega} t)} w^\mu\Big)
\nonumber
\\
& \times 
e^{-\frac{\Omega |u|^2 }{2\theta \tanh(\tilde{\Omega}t)}
-\frac{|w|^2}{2\Omega \theta \tanh(\tilde{\Omega}t)}
-2i\langle u,\Theta^{-1} z\rangle}\Big|_{w=u}\Big)
\nonumber
\\
&= 
\frac{1}{(2\pi\theta)^4} 
\int d^4u\,d^4 z  \,A_\mu(z)\Big(2\tilde{z}^\mu 
- \frac{(1-\Omega^2)}{\Omega \theta \tanh(\tilde{\Omega} t)}
\Theta^{\mu\nu}
\frac{\partial}{\partial z^\nu}\Big)
\frac{e^{-\frac{(1+\Omega^2) |u|^2 }{2\Omega \theta \tanh(\tilde{\Omega}t)}
-2i\langle u,\Theta^{-1} z\rangle}}{\sinh^4(\tilde{\Omega} t)} 
\nonumber
\\
&= 
\frac{\Omega^2}{(\pi\theta)^2(1+\Omega^2)^2} 
\int d^4 z  \,A_\mu(z)\Big(2\tilde{z}^\mu 
- \frac{(1-\Omega^2)}{\Omega \theta \tanh(\tilde{\Omega} t)}
\Theta^{\mu\nu}
\frac{\partial}{\partial z^\nu}\Big)
\frac{e^{-\frac{\tilde{\Omega} \tanh(\tilde{\Omega} t)}{(1+\Omega^2)}
    |z|^2 }}{\sinh^2(2\tilde{\Omega} t)} 
\nonumber
\\
&= 
\frac{4\Omega^4}{(\pi\theta)^2(1+\Omega^2)^3\sinh^2(2\tilde{\Omega} t)} 
\int d^4 z  \,\tilde{z}^\mu A_\mu(z)\, 
e^{-\frac{\tilde{\Omega} \tanh(\tilde{\Omega} t)}{1+\Omega^2} |z|^2 }\;.
\label{VDA}
\end{align}

This term gives the complete $A$-linear part. It vanishes for
$\Omega=0$, as expected.

\subsection{Two Moyal vertices}

To simplify the notations in this case we let 
$\tau_1 := \tanh(\tilde{\Omega} (t-t_2))$ and 
$\tau_2 := \tanh(\tilde{\Omega} t_2)$.
The change of variables $u_i=x_i-y_i$ and $v_i=x_i+y_i$ for $i=1,2$
leads to
\begin{align}
V_2(f,g)&:=\int d^4x_1 \,d^4y_1\, d^4x_2 \,d^4y_2\; 
(e^{-(t-t_2)H_0})(y_2,x_1) (L_\star(f))(x_1,y_1)
\nonumber
\\*
&\qquad\qquad \times (e^{-t_2H_0})(y_1,x_2) (L_\star(g))(x_2,y_2)
\nonumber
\\
&=  \Big(\frac{\tilde{\Omega}^2(1-\tau_1^2)(1-\tau_2^2)}{
16\pi^2\tau_1\tau_2}\Big)^2
\frac{1}{(2\pi\theta)^8}
\int d^4 u_1 \,d^4 v_1\, d^4 u_2 \,d^4v_2\,d^4z_1 \,d^4z_2\; 
f(z_1)g(z_2)
\nonumber
\\
& \times 
e^{
-\frac{\tilde{\Omega}}{16 \tau_1} |u_1+v_1+u_2-v_2|^2
-\frac{\tilde{\Omega}\tau_1}{16} |u_1+v_1-u_2+v_2|^2
-\frac{\tilde{\Omega}}{16 \tau_2} |u_1-v_1+u_2+v_2|^2
-\frac{\tilde{\Omega}\tau_2}{16} |-u_1+v_1+u_2+v_2|^2}
\nonumber
\\
& \times 
e^{i\langle u_1,\Theta^{-1} v_1\rangle
-2i\langle u_1,\Theta^{-1} z_1\rangle 
+ i\langle u_2,\Theta^{-1} v_2\rangle
-2i\langle u_2,\Theta^{-1} z_2\rangle }\;.
\end{align}
Defining 
\begin{align}
C &:=  \left(\begin{array}{cccc} 
1{+}\tau_1\tau_2 & 1{-}\tau_1\tau_2 &
-\frac{\tau_1-\tau_2}{\tau_1+\tau_2} (1{-}\tau_1\tau_2) &
\frac{\tau_1-\tau_2}{\tau_1+\tau_2} (1{+}\tau_1\tau_2) 
\\
1{-}\tau_1\tau_2 & 1{+}\tau_1\tau_2 &
-\frac{\tau_1-\tau_2}{\tau_1+\tau_2} (1{+}\tau_1\tau_2) &
\frac{\tau_1-\tau_2}{\tau_1+\tau_2} (1{-}\tau_1\tau_2) 
\\
-\frac{\tau_1-\tau_2}{\tau_1+\tau_2} (1{-}\tau_1\tau_2) &
-\frac{\tau_1-\tau_2}{\tau_1+\tau_2} (1{+}\tau_1\tau_2) &
1{+}\tau_1\tau_2 & -(1{-}\tau_1\tau_2) 
\\
\frac{\tau_1-\tau_2}{\tau_1+\tau_2} (1{+}\tau_1\tau_2) &
\frac{\tau_1-\tau_2}{\tau_1+\tau_2} (1{-}\tau_1\tau_2) &
-(1{-}\tau_1\tau_2) & 1{+}\tau_1\tau_2 
\end{array}\right)
\nonumber
\\
G &:= \left(\begin{array}{cccc}   
0 & 0 & \!\!\! -\frac{4\tau_1\tau_2}{\Omega(\tau_1+\tau_2)}\!\!\! & 0 \\
0 & 0 & 0 &\!\!\! -\frac{4\tau_1\tau_2}{\Omega(\tau_1+\tau_2)} \\
\frac{4\tau_1\tau_2}{\Omega(\tau_1+\tau_2)}\!\!\! & 0 & 0 & 0 \\  
0 & \!\!\!\frac{4\tau_1\tau_2}{\Omega(\tau_1+\tau_2)}\!\!\! & 0 & 0  \\  
\end{array}\right)\,,\quad
X:= \left(\begin{array}{c} 
u_1 \\ u_2 \\ v_1 \\ v_2
\end{array}\right),\quad 
Z:= \left(\begin{array}{c} 
z_1 \\ z_2 \\ 0 \\ 0
\end{array}\right),
\end{align}
and $Q:= C \otimes 1_4 + G \otimes \sigma$, we obtain 
\begin{align}
V_2(f,g)
&=  \Big(\frac{\tilde{\Omega}^2(1-\tau_1^2)(1-\tau_2^2)}{
16\pi^2\tau_1\tau_2}\Big)^2
\frac{1}{(2\pi\theta)^8}
\int d^{16} X d^8 Z\; 
f(z_1)g(z_2)
\nonumber
\\*
& \times 
e^{
-\frac{\Omega (\tau_1+\tau_2)}{8\theta \tau_1\tau_2} 
X^t Q X 
- \frac{2}{\theta} X^t \sigma Z }
\nonumber
\\
&=  \Big(\frac{\tilde{\Omega}^2(1-\tau_1^2)(1-\tau_2^2)}{
16\pi^2\tau_1\tau_2}\Big)^2
\Big(\frac{4\tau_1\tau_2}{\Omega (\tau_1+\tau_2)}\Big)^8
\det( C \otimes 1_4 + G \otimes \sigma)^{-\frac{1}{2}}
\nonumber
\\
& \times 
\int d^8 Z\; 
f(z_1)g(z_2)\;e^{
-\frac{8\tau_1\tau_2}{\Omega\theta (\tau_1+\tau_2)}
Z^t \sigma (C \otimes 1_4 + G \otimes \sigma)^{-1}  \sigma Z }\;.
\end{align}
In \cite{Gurau:2006yc} it was proven that
\begin{align}
\det (Q) &= \det (G+C)^4\;,\qquad \\
Q^{-1} &= 
\frac{1}{2} \big((G+C)^{-1} + ((G+C)^{-1})^t\big) \otimes 1_4
+  \frac{1}{2} \big((G+C)^{-1} - ((G+C)^{-1})^t\big) \otimes \sigma\;.
\end{align}
We find 
\begin{align}
&\det (G+C) = \Big(\frac{16(1+\Omega^2)
  \tau_1^2\tau_2^2}{\Omega^2 (\tau_1+\tau_2)^2}\Big)^2 
\Big(1 + \frac{\Omega^2(\tau_1-\tau_2)^2}{(1+\Omega^2)^2\tau_1\tau_2}\Big)\;,
\end{align}
which suggests to introduce 
\begin{align}
T:= 1 +
\frac{\Omega^2(\tau_1-\tau_2)^2}{(1+\Omega^2)^2\tau_1\tau_2}\;, 
\end{align}
and further
\begin{align}
&\frac{1}{2} \big((G+C)^{-1} + ((G+C)^{-1})^t\big)
=\frac{\Omega^2(\tau_1 + \tau_2)^2}{
16 \tau_1^2 \tau_2^2 (1 + \Omega^2)T}
\nonumber
\\
&\times \left(\!\! \begin{array}{cccc} 
1{+}\tau_1\tau_2 & -(1{-}\tau_1\tau_2) &
\frac{1-\Omega^2}{1+\Omega^2} 
\frac{\tau_1-\tau_2}{\tau_1+\tau_2} (1{-}\tau_1\tau_2) &
\frac{1-\Omega^2}{1+\Omega^2} 
\frac{\tau_1-\tau_2}{\tau_1+\tau_2} (1{+}\tau_1\tau_2) 
\\
-(1{-}\tau_1\tau_2) & 1{+}\tau_1\tau_2 &
-\frac{1-\Omega^2}{1+\Omega^2} 
\frac{\tau_1-\tau_2}{\tau_1+\tau_2} (1{+}\tau_1\tau_2) &
-\frac{1-\Omega^2}{1+\Omega^2} 
\frac{\tau_1-\tau_2}{\tau_1+\tau_2} (1{-}\tau_1\tau_2) 
\\
\frac{1-\Omega^2}{1+\Omega^2} 
\frac{\tau_1-\tau_2}{\tau_1+\tau_2} (1{-}\tau_1\tau_2) &
-\frac{1-\Omega^2}{1+\Omega^2} 
\frac{\tau_1-\tau_2}{\tau_1+\tau_2} (1{+}\tau_1\tau_2) &
1{+}\tau_1\tau_2 & 1{-}\tau_1\tau_2 
\\
\frac{1-\Omega^2}{1+\Omega^2} 
\frac{\tau_1-\tau_2}{\tau_1+\tau_2} (1{+}\tau_1\tau_2) &
-\frac{1-\Omega^2}{1+\Omega^2} 
\frac{\tau_1-\tau_2}{\tau_1+\tau_2} (1{-}\tau_1\tau_2) &
1{-}\tau_1\tau_2 & 1{+}\tau_1\tau_2 
\end{array}\!\! \right)
\end{align}
as well as 
\begin{align}
&\frac{1}{2} \big((G+C)^{-1} - ((G+C)^{-1})^t\big)
=\frac{\Omega^2(\tau_1 + \tau_2)^2}{
8 \tau_1^2 \tau_2^2 (1 + \Omega^2)^2T}
\nonumber
\\*
&\times \left(\begin{array}{cccc} 
0 & \Omega(\tau_1-\tau_2) &  \!\!\!\!\frac{2(1+\Omega^2)\tau_1\tau_2
  +\Omega^2(\tau_1-\tau_2)^2}{\Omega (\tau_1+\tau_2)} \!\!\!\! & 0 \\
 \!\!\!\!- \Omega(\tau_1-\tau_2) & 0 & 0  &\frac{2(1+\Omega^2)\tau_1\tau_2
  +\Omega^2(\tau_1-\tau_2)^2}{\Omega (\tau_1+\tau_2)} \!\!\!\! \\
- \frac{2(1+\Omega^2)\tau_1\tau_2+\Omega^2(\tau_1-\tau_2)^2}{
\Omega (\tau_1+\tau_2)} \!\!\!\! & 0 & 0 & 
\Omega(\tau_1-\tau_2) \\
0 &  \!\!\!\!- \frac{2(1+\Omega^2)\tau_1\tau_2+\Omega^2(\tau_1-\tau_2)^2}{
\Omega (\tau_1+\tau_2)}  \!\!\!\!& -\Omega(\tau_1-\tau_2) & 0
\end{array}\right).
\end{align}

We thus conclude 
\begin{align}
V_2(f,g) &= 
  \Big(\frac{\tilde{\Omega}^2(1-\tau_1^2)(1-\tau_2^2)}{
16(1+\Omega^2)^2 \pi^2 T  \tau_1\tau_2}\Big)^2
\int d^4z_1 d^4 z_2 \; f(z_1)g(z_2)\;
\nonumber
\\*
& \times 
e^{-\frac{\Omega(\tau_1 + \tau_2)}{2 \theta \tau_1 \tau_2 (1 +
\Omega^2)T}(|z_1-z_2|^2+\tau_1\tau_2|z_1+z_2|^2)
-\frac{2\Omega^2(\tau_1^2 - \tau_2^2)}{\theta
\tau_1 \tau_2 (1 + \Omega^2)^2T} z_1\sigma z_2}\;.
\end{align}
For $\tau_1,\tau_2 \to 0$ the integrand is regular unless
$z_1=z_2$. To capture the singularity at $z_1=z_2$, we expand 
$g(z_2)=g(z_1)+(z_2-z_1)\int d\xi \;(\partial_\mu g)(z_1+\xi(z_2-z_1))$
and consider the leading term $g(z_1)$. After a shift $z_2\mapsto
z_2+z_1$ we have
\begin{align}
V_2(f,g)^0 &= 
  \Big(\frac{\tilde{\Omega}(1-\tau_1^2)(1-\tau_2^2)}{
4\pi(1+\Omega^2)  (\tau_1+\tau_2)(1+\tau_1\tau_2)}\Big)^2
\int d^4z_1 \; f(z_1)g(z_1)\;
e^{-\frac{\tilde{\Omega}(\tau_1 + \tau_2)}{(1+ \tau_1 \tau_2) (1 +
\Omega^2)}|z_1|^2}\;.
\label{V2fg}
\end{align}
It can be shown that $(z_2-z_1)\int d\xi \;(\partial_\mu
g)(z_1+\xi(z_2-z_1))$ is subleading.

\subsection{Two Moyal-derivative vertices}

To complete the $A$-bilinear part, we also need the contribution with
two vertices of Moyal+derivative type. We use as far as possible the
same notation as in the previous calculation. Defining the auxiliary
vector $W=(0,0,w_3,w_4)^t$, this gives 
\begin{align}
  V_2(A,A)&:=\int d^4x_1 \,d^4y_1\, d^4x_2 \,d^4y_2\;
  (e^{-(t-t_2)H_0})(y_2,x_1) \big\{ L_\star(A^\mu) , i \partial_\mu +
  \Omega^2 M_\bullet(\tilde{x}_\mu) \big\}(x_1,y_1) \nonumber
  \\
  & \times (e^{-t_2H_0})(y_1,x_2) \big\{ L_\star(A^\mu) , i
  \partial_\mu + \Omega^2 M_\bullet(\tilde{x}_\mu) \big\}(x_2,y_2)
  \nonumber
  \\
&=
 \Big(\frac{\tilde{\Omega}^2(1-\tau_1^2)(1-\tau_2^2)}{
16\pi^2\tau_1\tau_2}\Big)^2
\frac{1}{(2\pi\theta)^8}
\int d^{16} X d^8 Z\; 
A_\mu(z_1)A_\nu(z_2)
\nonumber
\\
&\times 
 \Big(2 \tilde{z}_1^\mu  -(1-\Omega^2) (\Theta^{-1})^{\mu\rho} 
\frac{\partial}{\partial w_3^\rho}\Big) \Big(2
  \tilde{z}_2^\nu   -(1-\Omega^2) (\Theta^{-1})^{\nu\sigma} 
\frac{\partial}{\partial w_4^\sigma}\Big)
\nonumber
\\
& \times 
e^{-\frac{\Omega (\tau_1+\tau_2)}{8\theta \tau_1\tau_2} 
X^t Q X 
- \frac{2}{\theta} X^t \sigma Z +2 X^t W}\Big|_{W=0}
\nonumber
\\
&=
 \Big(\frac{\tilde{\Omega}^2(1-\tau_1^2)(1-\tau_2^2)}{
16\pi^2\tau_1\tau_2}\Big)^2
\Big(\frac{4 \tau_1\tau_2} {\Omega (\tau_1+\tau_2)}\Big)^8
(\det Q)^{-1/2}
\int d^8 Z\; A_\mu(z_1)A_\nu(z_2)
\nonumber
\\
&\times 
\Big(\frac{8\theta\tau_1\tau_2}{\Omega (\tau_1+\tau_2)}
(1-\Omega^2)^2 (\Theta^{-1})^{\mu\rho} 
(\Theta^{-1})^{\nu\sigma} \big((Q^{-1})^{43}_{\sigma\rho}+ 
(Q^{-1})^{34}_{\rho\sigma}\big) 
\nonumber
\\
&+
 \Big(2 \tilde{z}_1^\mu  +\frac{8i\theta\tau_1\tau_2(1-\Omega^2)}{
\Omega(\tau_1+\tau_2)} (\Theta^{-1})^{\mu\rho} 
(Q^{-1} \tilde{Z})_\rho^3\Big) 
\nonumber
\\
& \qquad \times \Big(2  \tilde{z}_2^\nu  
+\frac{8i\theta \tau_1\tau_2(1-\Omega^2)}{
\Omega(\tau_1+\tau_2)} (\Theta^{-1})^{\nu\sigma} 
(Q^{-1} \tilde{Z})_\sigma^4\Big) 
\Big)\nonumber
\\
& \times 
e^{-\frac{8\tau_1\tau_2}{\Omega\theta (\tau_1+\tau_2)}
(  Z(Q^{-1})Z + i\theta^2  W Q^{-1} \tilde{Z} -\theta^2 
W Q^{-1} W )}
\Big|_{W=0}
\nonumber
\\
&=
 \Big(\frac{\tilde{\Omega}^2(1-\tau_1^2)(1-\tau_2^2)}{
16(1+\Omega^2)^2\pi^2 T \tau_1\tau_2}\Big)^2
\int d^4 z_1 \,d^4z_2\; A_\mu(z_1)A_\nu(z_2)
\nonumber
\\
&\times 
\Big(\frac{\Omega(1-\Omega^2)^2 
(\tau_1 + \tau_2)(1-\tau_1\tau_2)}{\theta
(1 + \Omega^2)T\tau_1 \tau_2 }
\delta^{\mu\nu}
\nonumber
\\
& 
+ \Big(  -
\frac{i\tilde{\Omega}(\tau_1 - \tau_2)}{2\tau_1 \tau_2T}
\frac{(1-\Omega^2)^2}{(1+\Omega^2)^2} 
\big((z_1^\mu-z_2^\mu)
- \tau_1\tau_2(z_1^\mu +z_2^\mu)\big) 
+\frac{\Omega^2(\tau_1+\tau_2)^2}{
(1+\Omega^2)\tau_1\tau_2 T} \tilde{z}_1^\mu
\Big)
\nonumber
\\
& \times \Big(
-
\frac{i\tilde{\Omega}(\tau_1 - \tau_2)}{2\tau_1 \tau_2 T}
\frac{(1-\Omega^2)^2}{(1+\Omega^2)^2} 
\big((z_1^\nu-z_2^\nu)+\tau_1\tau_2(z_1^\nu+z_2^\nu)\big)
+ \frac{\Omega^2(\tau_1+\tau_2)^2}{
(1+\Omega^2)\tau_1\tau_2 T} \tilde{z}_2^\nu\Big)\Big)
\nonumber
\\
& \times e^{-\frac{\Omega(\tau_1 + \tau_2)}{2 \theta \tau_1 \tau_2 (1 +
\Omega^2)T}(|z_1-z_2|^2+\tau_1\tau_2|z_1+z_2|^2)
-\frac{2\Omega^2(\tau_1^2 - \tau_2^2)}{\theta
\tau_1 \tau_2 (1 + \Omega^2)^2T} z_1\sigma z_2}\;.
\end{align}
We write $(z_1-z_2) \pm\tau_1\tau_2(z_1+z_2)$ as derivative of the
exponential, plus appropriate corrections, and integrate by parts:
\begin{align}
V_2(A,A)&=
 \Big(\frac{\tilde{\Omega}^2(1-\tau_1^2)(1-\tau_2^2)}{
16(1+\Omega^2)^3\pi^2 T \tau_1\tau_2}\Big)^2
\int d^4 z_1 \,d^4z_2\; A_\mu(z_1)A_\nu(z_2)
\nonumber
\\
&\times 
\Big(
\frac{2\tilde{\Omega}(1-\Omega^2)^2(1+\Omega^2) (1-\tau_1\tau_2)}{
(\tau_1+\tau_2)}\delta_{\mu\nu}
- 2i \Omega^2(1-\Omega^2)^2
\frac{(\tau_1^2-\tau_2^2)}{\tau_1\tau_2 T} (\Theta^{-1})^{\mu\nu} 
\nonumber
\\
& 
+ \Big(  -i(1-\Omega^2)^2
\frac{\tau_1-\tau_2}{\tau_1+\tau_2} \frac{\partial}{\partial z_{2\mu}}
+4\Omega^2 \tilde{z}_1^\mu\Big)
\Big(
i(1-\Omega^2)^2
\frac{\tau_1-\tau_2}{\tau_1+\tau_2}  \frac{\partial}{\partial z_{1\nu}}
+ 4\Omega^2\tilde{z}_2^\nu\Big)\Big)
\nonumber
\\
& \times e^{-\frac{\Omega(\tau_1 + \tau_2)}{2 \theta \tau_1 \tau_2 (1 +
\Omega^2)T}(|z_1-z_2|^2+\tau_1\tau_2|z_1+z_2|^2)
-\frac{2\Omega^2(\tau_1^2 - \tau_2^2)}{\theta
\tau_1 \tau_2 (1 + \Omega^2)^2T} z_1\sigma z_2}
\nonumber
\\
&= \Big(\frac{\tilde{\Omega}^2(1-\tau_1^2)(1-\tau_2^2)}{
16(1+\Omega^2)^3\pi^2 T \tau_1\tau_2}\Big)^2
\int d^4 z_1 \,d^4z_2\; 
\nonumber
\\
&
\times \Big( \frac{2\tilde{\Omega}(1-\Omega^2)^2(1+\Omega^2) 
(1-\tau_1\tau_2)}{\tau_1+\tau_2} A_\mu(z_1)A^\mu(z_2)
+16\Omega^4 \tilde{z}_1^\mu  A_\mu(z_1)\,\tilde{z}_2^\nu A_\nu(z_2)
\nonumber
\\
& + (1-\Omega^2)^4 \frac{(\tau_1-\tau_2)^2}{(\tau_1+\tau_2)^2} 
(\partial_\nu A_\mu)(z_1)\,(\partial_\mu A_\nu)(z_2)
\nonumber
\\
&+ 4i \Omega^2(1-\Omega^2)^2
\frac{\tau_1-\tau_2}{\tau_1+\tau_2} 
\big( 
A^\mu(z_1) \tilde{z}_2^\nu (\partial_\mu A_\nu)(z_2)
-\tilde{z}_1^\mu (\partial_\nu A_\nu)(z_1)\,A^\nu(z_2) \big)
\nonumber
\\
&-  2i \Omega^2(1-\Omega^2)^2
\frac{\tau_1-\tau_2}{\tau_1+\tau_2}
\Big(4+ \frac{(\tau_1+\tau_2)^2}{\tau_1\tau_2 T}\Big)
(\Theta^{-1})^{\mu\nu} A_\mu(z_1) A_\nu(z_2)
\Big)\nonumber
\\
& \times e^{-\frac{\tilde{\Omega}(\tau_1 + \tau_2)}{4 \tau_1 \tau_2 (1 +
\Omega^2)T}(|z_1-z_2|^2+\tau_1\tau_2|z_1+z_2|^2)
-\frac{2i\Omega^2(\tau_1^2 - \tau_2^2)}{
\tau_1 \tau_2 (1 + \Omega^2)^2T} \langle z_1,\Theta^{-1} z_2\rangle}\;.
\end{align}
Again, the integrand is regular for $z_1\neq z_2$, so that we expand 
\begin{align}
A_\nu(z_2) &= A_\nu(z_1)
+(z_2^\rho-z_1^\rho)(\partial_\rho A_\nu)(z_1)
+\frac{1}{2}(z_2^\rho-z_1^\rho)(z_2^\sigma-z_1^\sigma)
(\partial_\rho\partial_\sigma A_\nu)(z_1) 
\nonumber
\\
& + \frac{1}{2}(z_2^\rho{-}z_1^\rho)(z_2^\sigma{-}z_1^\sigma)
(z_2^\kappa{-}z_1^\kappa)\int_0^1 d\xi\;(1{-}\xi)^2
(\partial_\rho\partial_\sigma\partial_\kappa A_\nu)(z_1{+}\xi(z_2{-}z_1)) \;,
\end{align}
and similarly for $(\partial_\mu A_\nu)(z_2)$. In leading $t$-order,
we must expand $A_\mu(z_1)A^\mu(z_2)$ up to second order (due to the
appearance of $(\tau_1+\tau_2)^{-1}$) and all other terms only up to
zeroth order. These leading terms become after a shift $z_2\mapsto
z_2+z_1$
\begin{align}
V_2(A,A)^0 &= \Big(\frac{\tilde{\Omega}^2(1-\tau_1^2)(1-\tau_2^2)}{
16(1+\Omega^2)^3\pi^2 T \tau_1\tau_2}\Big)^2
\int d^4 z_1 \,d^4z_2\; 
\nonumber
\\
&
\times \Big( \frac{2\tilde{\Omega}(1-\Omega^2)^2(1+\Omega^2) 
(1-\tau_1\tau_2)}{\tau_1+\tau_2} \Big(
A_\mu(z_1)A^\mu(z_1)+ A_\mu(z_1)(\partial_\rho A^\mu)(z_1)
\frac{\partial}{\partial w_\rho}
\nonumber
\\ &\qquad + 
\frac{1}{2} A_\mu(z_1)(\partial_\rho\partial_\sigma  A^\mu)(z_1)
\frac{\partial^2}{\partial w_\rho \partial w_\sigma}\Big)
\nonumber
\\
& +16\Omega^4 \tilde{z}_1^\mu  A_\mu(z_1)\,
\tilde{z}_1^\nu A_\nu(z_1)
+ (1-\Omega^2)^4 \frac{(\tau_1-\tau_2)^2}{(\tau_1+\tau_2)^2} 
(\partial_\nu A_\mu)(z_1)\,(\partial_\mu A_\nu)(z_1)
\nonumber
\\
&+ 2(\Theta^{-1})^{\nu\rho} \Big( 4i \Omega^2(1{-}\Omega^2)^2
\frac{\tau_1-\tau_2}{\tau_1+\tau_2} 
A^\mu(z_1) (\partial_\mu A_\nu)(z_1)
+16\Omega^4 \tilde{z}_1^\mu  A_\mu(z_1)\,
A^\nu(z_1)\Big) \frac{\partial}{\partial
  w^\rho}\Big)\nonumber
\\
& \times e^{-\frac{\tilde{\Omega}(\tau_1 + \tau_2)}{4 \tau_1 \tau_2 (1 +
\Omega^2)T}(1+\tau_1\tau_2)|z_2|^2+4\tau_1\tau_2 \langle
z_2,z_1\rangle +4\tau_1\tau_2 |z_1|^2)
+\frac{2i\Omega^2(\tau_1^2 - \tau_2^2)}{
\tau_1 \tau_2 (1 + \Omega^2)^2T} \langle z_2,\Theta^{-1} z_1\rangle
+\langle w,z_2\rangle}\Big|_{w=0}
\nonumber
\\
 &= \Big(\frac{\tilde{\Omega}(1-\tau_1^2)(1-\tau_2^2)}{
4\pi(1+\Omega^2)^2 ( \tau_1+\tau_2)(1+\tau_1\tau_2)}\Big)^2
\int d^4 z_1 \; 
e^{-\frac{\tilde{\Omega}(\tau_1 + \tau_2)}{
(1 + \Omega^2)(1+\tau_1\tau_2) } |z_1|^2}
\nonumber
\\
&
\times \Big( \frac{2\tilde{\Omega}(1-\Omega^2)^2(1+\Omega^2) 
(1-\tau_1\tau_2)}{\tau_1+\tau_2} \Big(
A_\mu(z_1)A^\mu(z_1)
\nonumber
\\
& \qquad + A^\mu(z_1)(\partial_\nu A_\mu)(z_1)
\Big(-\frac{2\tau_1\tau_2}{1+\tau_1\tau_2} z_1^\nu
+\frac{i\theta \Omega (\tau_1-\tau_2)}{(1+\Omega^2)(1+\tau_1\tau_2)}
\tilde{z}_1^\nu\Big)
\nonumber
\\ &\qquad 
+ 
\frac{A_\mu(z_1)(\partial_\rho\partial_\sigma  A^\mu)(z_1)}{
2(1+\tau_1\tau_2)^2} 
\Big(2\tau_1\tau_2 z_1^\rho
-\frac{i\theta \Omega (\tau_1-\tau_2)}{(1+\Omega^2)}
\tilde{z}_1^\rho\Big)
\Big(2\tau_1\tau_2 z_1^\sigma
-\frac{i\theta \Omega (\tau_1-\tau_2)}{(1+\Omega^2)}
\tilde{z}_1^\sigma\Big)
\nonumber
\\
&\qquad 
+ A^\mu(z_1)(\partial^\nu\partial_\nu  A_\mu)(z_1)
\frac{\tau_1 \tau_2 (1 +\Omega^2)T}{
\tilde{\Omega}(\tau_1 + \tau_2)(1+\tau_1\tau_2)} \Big)
\nonumber
\\
& +16\Omega^4 \tilde{z}_1^\mu  A_\mu(z_1)\,
\tilde{z}_1^\nu A_\nu(z_1)
+ (1-\Omega^2)^4 \frac{(\tau_1-\tau_2)^2}{(\tau_1+\tau_2)^2} 
(\partial_\nu A_\mu)(z_1)\,(\partial_\mu A_\nu)(z_1)
\nonumber
\\
&+  \Big( 4i \Omega^2(1-\Omega^2)^2
\frac{\tau_1-\tau_2}{\tau_1+\tau_2} 
A^\mu(z_1) (\partial_\mu A_\nu)(z_1)
+16\Omega^4 \tilde{z}_1^\mu  A_\mu(z_1)\,
A_\nu(z_1)\Big) 
\nonumber
\\
&\qquad\times \Big(
-\frac{2\tau_1\tau_2}{1+\tau_1\tau_2} \tilde{z}_1^\nu
-\frac{2i \tilde{\Omega} (\tau_1-\tau_2)}{(1+\Omega^2)(1+\tau_1\tau_2)}
z_1^\nu\Big)\Big)\;.
\label{V2A0}
\end{align}

\subsection{Moyal vertex plus Moyal-derivative vertex}

This combination is (among others) necessary for a new type of
coupling between scalar field and gauge field. We use as far as possible the
same notation as in the previous calculation. Defining the auxiliary
vector $W=(0,0,w_3,w_4)^t$, we have
\begin{align}
  V_2(A,f)&=\int d^4x_1 \,d^4y_1\, d^4x_2 \,d^4y_2\;
  (e^{-(t-t_2)H_0})(y_2,x_1) (L_\star(f))(x_1,y_1)
\nonumber
  \\
  & \times (e^{-t_2H_0})(y_1,x_2) 
\big\{ L_\star(A^\mu) , i \partial_\mu +
  \Omega^2 M_\bullet(\tilde{x}_\mu) \big\}(x_1,y_1) 
  \nonumber
  \\
&=
 \Big(\frac{\tilde{\Omega}^2(1-\tau_1^2)(1-\tau_2^2)}{
16\pi^2\tau_1\tau_2}\Big)^2
\frac{1}{(2\pi\theta)^8}
\int d^{16} X d^8 Z\; 
f(z_1)A_\mu(z_2)
\nonumber
\\
&\times 
 \Big(2 \tilde{z}_2^\mu  -(1-\Omega^2) (\Theta^{-1})^{\mu\rho} 
\frac{\partial}{\partial w_4^\rho}\Big) 
e^{-\frac{\Omega (\tau_1+\tau_2)}{8\theta \tau_1\tau_2} 
X^t Q X 
- \frac{2}{\theta} X^t \sigma Z +2 X^t W}\Big|_{W=0}
\nonumber
\\
&=
 \Big(\frac{\tilde{\Omega}^2(1-\tau_1^2)(1-\tau_2^2)}{
16\pi^2\tau_1\tau_2}\Big)^2
\Big(\frac{4 \tau_1\tau_2} {\Omega (\tau_1+\tau_2)}\Big)^8
(\det Q)^{-1/2}
\int d^8 Z\; f(z_1)A_\mu(z_2)
\nonumber
\\
&\times 
 \Big(2 \tilde{z}_2^\mu  +\frac{8i\theta\tau_1\tau_2(1-\Omega^2)}{
\Omega(\tau_1+\tau_2)} (\Theta^{-1})^{\mu\rho} 
(Q^{-1} \tilde{Z})_\rho^4\Big) 
e^{-\frac{8\tau_1\tau_2}{\Omega\theta (\tau_1+\tau_2)}
Z(Q^{-1})Z }
\nonumber
\\
&=
 \Big(\frac{\tilde{\Omega}^2(1-\tau_1^2)(1-\tau_2^2)}{
16(1+\Omega^2)^2\pi^2 T \tau_1\tau_2}\Big)^2
\int d^4 z_1 \,d^4z_2\; f(z_1)A_\mu(z_2)
\nonumber
\\
& \times \Big(
-\frac{i\tilde{\Omega}(\tau_1 - \tau_2)}{2\tau_1 \tau_2 T}
\frac{(1-\Omega^2)^2}{(1+\Omega^2)^2} 
\big((z_1^\mu-z_2^\mu)+\tau_1\tau_2(z_1^\mu+z_2^\mu)\big)
+ \frac{\Omega^2(\tau_1+\tau_2)^2}{
(1+\Omega^2)\tau_1\tau_2 T} \tilde{z}_2^\mu\Big)
\nonumber
\\
& \times e^{-\frac{\Omega(\tau_1 + \tau_2)}{2 \theta \tau_1 \tau_2 (1 +
\Omega^2)T}(|z_1-z_2|^2+\tau_1\tau_2|z_1+z_2|^2)
-\frac{2\Omega^2(\tau_1^2 - \tau_2^2)}{\theta
\tau_1 \tau_2 (1 + \Omega^2)^2T} z_1\sigma z_2}
\nonumber
\\
&= \Big(\frac{\tilde{\Omega}^2(1-\tau_1^2)(1-\tau_2^2)}{
16(1+\Omega^2)^2\pi^2 T \tau_1\tau_2}\Big)^2
\int d^4 z_1 \,d^4z_2\; 
\nonumber
\\
& 
\times \Big( -i\frac{(1-\Omega^2)^2}{1+\Omega^2}
\frac{\tau_1-\tau_2}{\tau_1+\tau_2}  (\partial^\mu f)(z_1)
\,A_\mu(z_2) + \frac{4\Omega^2}{1+\Omega^2} \;
f(z_1)\, \tilde{z}_2^\mu A_\mu(z_2)\Big)
\nonumber
\\
& \times e^{-\frac{\Omega(\tau_1 + \tau_2)}{2 \theta \tau_1 \tau_2 (1 +
\Omega^2)T}(|z_1-z_2|^2+\tau_1\tau_2|z_1+z_2|^2)
-\frac{2\Omega^2(\tau_1^2 - \tau_2^2)}{\theta
\tau_1 \tau_2 (1 + \Omega^2)^2T} z_1\sigma z_2}\;.
\end{align}
As before, for $\tau_1,\tau_2 \to 0$ the integrand is regular unless
$z_1=z_2$, so that we expand 
$A_\mu(z_2)=A_\mu(z_1)+(z_2^\nu-z_1^\nu)\int d\xi \;(\partial_\nu
A_\mu)(z_1+\xi(z_2-z_1))$
and consider the leading term $A_\mu(z_1)$. After a shift $z_2\mapsto
z_2+z_1$ we have, neglecting the subleading summand $\tilde{z}_2^\mu$,
\begin{align}
V_2(A,f)^0 &=  \Big(\frac{\tilde{\Omega}(1-\tau_1^2)(1-\tau_2^2)}{
4\pi(1+\Omega^2) (\tau_1+\tau_2)(1+\tau_1\tau_2)}\Big)^2
\int d^4 z_1 \; 
e^{-\frac{\tilde{\Omega}(\tau_1 + \tau_2)}{(1 +
\Omega^2)(1+\tau_1\tau_2)}|z_1|^2}
\nonumber
\\
& 
\times \Big( i\frac{(1-\Omega^2)^2}{1+\Omega^2}
\frac{\tau_1-\tau_2}{\tau_1+\tau_2}  f(z_1)
\,(\partial^\mu A_\mu)(z_1) + \frac{4\Omega^2}{1+\Omega^2} \;
f(z_1)\, \tilde{z}_1^\mu A_\mu(z_1)\Big)\;.
\label{VDAf}
\end{align}

\end{document}